\documentclass[floats,floatfix,amssymb,prd,twocolumn,superscriptaddress,nofootinbib]{revtex4-1}
	
\usepackage{subcaption}
\usepackage{ragged2e}
\DeclareCaptionJustification{justified}{\justifying}
\usepackage{amssymb,amsmath,verbatim,mathtools,needspace,enumitem,etoolbox,graphicx,physics,microtype,afterpage,bm}
\usepackage{esint}
\usepackage[dvipsnames, usenames]{xcolor}
\definecolor{linkcolor}{rgb}{0.0,0.3,0.5}
\usepackage[unicode, colorlinks=true, linkcolor=linkcolor, citecolor=linkcolor, filecolor=linkcolor,urlcolor=linkcolor, pdfusetitle]{hyperref}
\usepackage[all]{hypcap}
\usepackage[T1]{fontenc}
\usepackage[utf8]{inputenc}
\usepackage{tabularx}
\usepackage{float}
\newcommand\underrel[3][]{\mathrel{\mathop{#3}\limits_{%
      \ifx c#1\relax\mathclap{#2}\else#2\fi}}}

\usepackage{soul}

\usepackage{lmodern}
\allowdisplaybreaks
\usepackage{tikz}
\usepackage{color}
\usepackage{framed}
\usepackage{hyperref}
\hypersetup{colorlinks, citecolor=bluscuro, linkcolor=black, urlcolor=bluscuro}
\definecolor{rossos}{cmyk}{0,1,1,0.55}
\definecolor{bluscuro}{rgb}{0.15, 0.2, .85}
\definecolor{bluchiaro}{cmyk}{1,.3,0.,0.1}
\definecolor{ForestGreen}{rgb}{0.13, 0.55, 0.13}

\newcommand{\be}{\begin{equation}}
\newcommand{\ee}{\end{equation}}
\renewcommand{\d}{{\rm d}}

\newcommand{\llp}{\left [}
\newcommand{\rrp}{\right ]}
\newcommand{\lp}{\left (}
\newcommand{\rp}{\right )}

\def\lsim{\mathrel{\rlap{\lower4pt\hbox{\hskip0.5pt$\sim$}}
    \raise1pt\hbox{$<$}}}         
\def\gsim{\mathrel{\rlap{\lower4pt\hbox{\hskip0.5pt$\sim$}}
    \raise1pt\hbox{$>$}}}         

\makeatletter
\newcommand{\subsetsim}{\mathrel{\mathpalette\subset@sim\relax}}
\newcommand{\subset@sim}[2]{%
  \vtop{\offinterlineskip\m@th
    \ialign{\hfil##\cr
     ~$#1\subset$\cr\noalign{\kern0.5pt}\scalebox{0.9}{$#1\sim$}\cr
    }%
  }%
}
\makeatother

\makeatletter
\def\l@subsubsection#1#2{}
\makeatother

\begin{document}

\title{Tidal Love numbers and Green's functions in black hole spacetimes}

\author{Valerio De Luca}
\email{vdeluca@sas.upenn.edu}
\affiliation{Center for Particle Cosmology, Department of Physics and Astronomy,
University of Pennsylvania 209 South 33rd Street, Philadelphia, Pennsylvania 19104, USA}

\author{Alice Garoffolo}
\email{aligaro@sas.upenn.edu}
\affiliation{Center for Particle Cosmology, Department of Physics and Astronomy,
University of Pennsylvania 209 South 33rd Street, Philadelphia, Pennsylvania 19104, USA}

\author{Justin Khoury}
\email{jkhoury@upenn.edu}
\affiliation{Center for Particle Cosmology, Department of Physics and Astronomy,
University of Pennsylvania 209 South 33rd Street, Philadelphia, Pennsylvania 19104, USA}

\author{Mark Trodden}
\email{trodden@upenn.edu}
\affiliation{Center for Particle Cosmology, Department of Physics and Astronomy,
University of Pennsylvania 209 South 33rd Street, Philadelphia, Pennsylvania 19104, USA}


\begin{abstract}
\noindent
Tidal interactions play a crucial role in deciphering gravitational wave signals emitted by the coalescence of binary systems. They are usually quantified by a set of complex coefficients which include tidal Love numbers, describing the conservative response to an external perturbation. In the static case, these are found to vanish exactly for asymptotically flat black holes in general relativity in four spacetime dimensions, and recently they have been generalized to dynamical interactions. In the context of response theory, the retarded Green's function provides the complete description of the behavior of dynamical systems.
In this work we investigate the relation between Love numbers and Green's functions, and highlight the relevance of radiation reaction effects to their connection. As a special case, we discuss Banados-Teitelboim-Zanelli black holes, where the absence of radiative modes allows us to make a direct link between them.
\end{abstract}

\maketitle

\section{Introduction}
\label{sec:intro}
\noindent
The detection of gravitational waves emitted during the merger of binary systems marks a groundbreaking era in the study of gravity and compact objects in the strong-field regime~\cite{Bailes:2021tot}. Since the number of observations is expected to increase significantly when the next generation of detectors are operational~\cite{Punturo:2010zz, Sathyaprakash:2019yqt, Maggiore:2019uih, Reitze:2019iox,  Kalogera:2021bya}, it is necessary to have highly accurate waveform templates based on a precise understanding of the conservative and dissipative dynamics of two-body systems~\cite{Chia:2023tle, Chia:2024bwc}. In this regard, a crucial aspect of the inspiral of binary systems is tidal effects. Their measurement with gravitational wave observations allows us to glean valuable insights into the interior structure of compact objects. For example, they could reveal information about the equations of state of neutron stars (see Refs.~\cite{GuerraChaves:2019foa,Chatziioannou:2020pqz} for reviews), shed light on the presence of exotic compact objects~\cite{Cardoso:2017cfl, Cardoso:2019rvt, Herdeiro:2020kba,Chen:2023vet, Berti:2024moe}, and potentially unveil new aspects of physics at the event horizons of black holes (BHs)~\cite{Maselli:2018fay, Datta:2021hvm}.

An enduring analytical framework to characterize tidal effects is a set of complex coefficients. Their real parts, often called tidal Love numbers (TLNs), describe the linear conservative response of self-gravitating bodies~\cite{1909MNRAS..69..476L}; their imaginary parts are instead associated to dissipative effects. Initially formulated within the realm of Newtonian gravity, TLNs have undergone further developments to encompass a fully relativistic context~\cite{Hinderer:2007mb,Binnington:2009bb,Damour:2009vw}.
Within this framework, it is widely recognized that the static TLNs of different families of asymptotically flat BHs in four-dimensional spacetime, such as Schwarzschild, Kerr and Reissner-Nordstrom BHs, are precisely zero~\cite{Binnington:2009bb,Damour:2009vw,Damour:2009va,Pani:2015hfa,Pani:2015nua,Gurlebeck:2015xpa,Porto:2016zng,LeTiec:2020spy, Chia:2020yla,LeTiec:2020bos,Charalambous:2021mea,Charalambous:2021kcz,Bonelli:2021uvf,Ivanov:2022hlo,Charalambous:2022rre,Katagiri:2022vyz, Ivanov:2022qqt,Berens:2022ebl, Bhatt:2023zsy, Sharma:2024hlz, Rai:2024lho}. 
However, this characteristic is fragile, as it is violated in scenarios involving BH mimickers~\cite{Pani:2015tga,Cardoso:2017cfl}, in the presence of a cosmological constant~\cite{Nair:2024mya} or extended gravitational theories~\cite{Cardoso:2017cfl,Cardoso:2018ptl,DeLuca:2022tkm, Barura:2024uog}, in higher dimensions~\cite{Kol:2011vg,Cardoso:2019vof, Hui:2020xxx,Rodriguez:2023xjd,Charalambous:2023jgq,Charalambous:2024tdj, Charalambous:2024gpf}, or in nonvacuum environments induced by secular effects such as accretion or superradiant instabilities of ultralight bosonic fields~\cite{Baumann:2018vus,Cardoso:2019upw,DeLuca:2021ite,Cardoso:2021wlq, DeLuca:2022xlz}. 

Recently, there has been an emerging interest in incorporating nonlinear effects in TLNs~\cite{DeLuca:2023mio, Riva:2023rcm, Ivanov:2024sds}, as well as going beyond the assumption of static perturbations to compute dynamical Love numbers~\cite{Nair:2022xfm, Saketh:2023bul, Perry:2023wmm, Chakraborty:2023zed}. Those describe the conservative response to an external time-dependent, low-frequency perturbation, and can be thought of as a correction to the static coefficient due to dynamical processes. While static TLNs appear in the gravitational waveform at the 5th post-Newtonian (PN) order~\cite{Flanagan:2007ix}, the dynamical TLNs affect it at the 8th PN order~\cite{Saketh:2023bul}. Although they arise at higher PN order, excluding the dynamical Love numbers from a waveform model could introduce a bias in the measurements~\cite{Pratten:2021pro}. For neutron stars, their values have been constrained with gravitational data from the event GW170817~\cite{PhysRevD.100.021501, 2020NatCo..11.2553P}.

The whole set of TLNs and dissipative coefficients are usually defined within the context of {\it response theory}, which provides a theoretical framework 
to describe how physical systems react to external forces or perturbations~\cite{denittis2016linear}. For example, in electromagnetism, response theory can be used to understand how materials respond to electric and magnetic fields, leading to phenomena like polarization and magnetization. In response theory, the linear reaction of a system to an external force is usually described by the retarded Green's function. Indeed, by properly convolving the Green's function with the external perturbation, one can determine the system's response at any point in space and time. It is therefore natural to inquire whether there is a direct relation between the tidal response, parameterized by TLNs and dissipative coefficients, and the system's Green's function. While such a connection seems logical, it is a subtle problem to properly disentangle the various contributions entering the Green's function.

Among these contributions, one of the most significant is radiation reaction.
The concept of radiation reaction in gravity plays a crucial role in the study of coalescing binary systems (see Refs.~\cite{Detweiler:2005kq, Poisson:2011nh, Barack:2018yvs} for reviews). There, the energy and momentum carried away by the emitted gravitational waves result in the binary constituents experiencing a reaction force, which influences their motion and changes their trajectory and energy. This effect is completely analogous to the radiation reaction in classical electrodynamics, leading to the familiar Abraham-Lorentz force. The most striking application concerns extreme mass-ratio inspirals, describing the evolution of a binary system with a large hierarchy in masses between its constituents. At leading order in the mass ratio, one has a pointlike particle moving in a geodesic orbit around the heavier BH, while at subsequent orders the interaction of the particle with its own gravitational perturbation gives rise to an effective “self-force”, which changes the evolution of the orbit.

The purpose of this paper is to point out that both the tidal response, captured by TLNs and dissipative coefficients, and radiation reaction effects enter the full response of a self-gravitating system to an external perturbation. We will show that, while the static response is free from this contamination, the dynamical response to an external time-dependent force may be overshadowed by radiation reaction phenomena, such as tail effects. 

The outline of the paper is the following. In Sec.~\ref{sec: GF} we discuss the  retarded Green's function in linear response theory. Its connection to TLNs, as well as a review on the topic, is performed in Sec.~\ref{sec: TLN} for Newtonian gravity. In Sec.~\ref{sec: TLN-GR} we overview the formalism to extract TLNs in general relativity, which we then apply to Schwarzschild BHs. In Sec.~\ref{sec: TLN&tail} we point out a connection between tidal response and tail effects, discussing the special case of Banados-Teitelboim-Zanelli (BTZ) BHs in Sec.~\ref{sec: BTZ}. The conclusions are left to Sec.~\ref{conclusions}. 
In this paper we use natural units~$\hbar = c = 1$, and mostly positive signature.

\section{Retarded Green's function in linear response theory}
\label{sec: GF}
\noindent
In this Section we describe some of the properties of the retarded Green's function, describing the propagation of small linear perturbations on BH backgrounds. For simplicity, we focus on a free scalar field on a fixed BH background. Since we are interested in TLNs, the scalar is assumed massless. To keep the discussion as general as possible, we intentionally keep the notation schematic in this section. We will be more precise in specific examples later on in the paper.

A massless scalar field on a BH background satisfies the wave equation
\begin{equation}
\label{boxphi}
\Box \Phi = 0\,.
\end{equation}
Focusing on a single frequency mode~$\omega$, and performing a multipole moment decomposition,
the wave equation gives rise to a Schrödinger-like radial equation
\begin{equation}
\label{Schrodinger}
\frac{\d^2 \tilde{R}}{\d x^2}  + \left(\omega^2 - V(x)\right) \tilde{R} = 0\,,
\end{equation}
where~$x$ is the tortoise coordinate for the spacetime of interest,~$\tilde R (x)$ is the (rescaled) radial part of the field profile,
and~$V$ is an effective potential encoding the information of the background geometry. Examples of wavelike equations of this form are those for odd (Regge-Wheeler)~\cite{Regge:1957td} and even (Zerilli)~\cite{Zerilli:1970se} gravitational perturbations on Schwarzschild BHs, or the Teukolsky equations for perturbations on a Kerr geometry~\cite{Teukolsky:1973ha}. Later we will explicitly consider equations of this type for Kerr and BTZ BHs.

In response theory, compact sources perturbing the BH induce a response in the system inherited by the function~$\Phi$. This is manifest when one studies gravitational perturbations on a BH background, in which the metric perturbation encapsulates the deformation of the BH. In order to understand the time evolution of this perturbation,  one needs to specify the Cauchy data at some initial time. For general source terms, causality requires~$\Phi (x,t)$ and~$\partial_t \Phi (x,t)$ to vanish for all times prior to the appearance of the source (which we set at the time~$t_1 = 0$)~\cite{Leaver:1986gd}. This problem is typically approached via the retarded Green's function, defined by the equation
\begin{equation}
\label{eq:EQ_Green_function}
\left( \frac{\d^2 }{\d x^2} - \frac{\d^2 }{\d t^2} -  V \right) G(x_1,t_1; x_2, t_2) = \delta(t_2-t_1) \delta (x_2 - x_1)\,.
\end{equation}
The above equation is supplemented by the requirement of vanishing~$G$ whenever~$(x_2,t_2)$ lies outside the causal future of~$(x_1,t_1)$. Employing the initial data, one can perform a Fourier transform to define
\begin{align}
\label{Greentransform}
G (x_1,x_2;\omega) & = \int_0^{+\infty} \d t_2 \, {\rm e}^{{\rm i} \omega t_2} \, G(x_1,0; x_2, t_2)\,; \nonumber \\
G(x_1,0; x_2, t_2) & = \frac{1}{2\pi} \int_{-\infty + {\rm i}c}^{+\infty + {\rm i}c} \d \omega \, {\rm e}^{-{\rm i} \omega t_2}\, G (x_1,x_2;\omega)\,,
\end{align}
where~$c$ is some small positive number, and where the transform is well defined as long as~${\rm Im}\, \omega \geq 0$~\cite{PhysRevD.51.353}. The Green's function~$G (x_1,x_2;\omega)$ can be expressed in terms of two independent solutions of the homogeneous equation~\eqref{Schrodinger} and demands the fixing of proper boundary conditions for the response problem. These conditions are defined by inspecting the asymptotic behaviors of the solutions, and amount to considering a wave coming in from infinity, scattering off the potential, and being partly reflected and partly absorbed by the BH.

For asymptotically flat spacetimes, the first solution~$\tilde{R}^+$ is defined by imposing purely ingoing waves crossing the event horizon~$x_+$,~$\tilde{R}^+ (x) \sim {\rm e}^{-{\rm i} \omega x}~$ as~$x \to x_+$, while the second solution~$\tilde{R}^\infty$ is obtained by imposing purely outgoing waves near infinity~$x_\infty$ ({\it i.e.}, no initial incoming wave from infinity is allowed),~$\tilde{R}^\infty (x) \sim {\rm e}^{+{\rm i} \omega x}~$ as~$x \to x_\infty$. The Green's function constructed out of them is then~\cite{Leaver:1986gd}
\begin{align}
\label{eq:GreenFunction_Definition}
G (x_1,x_2;\omega) = \frac{ \tilde{R}^+ (x_1) \tilde{R}^\infty (x_2) }{W \left[\tilde{R}^\infty, \tilde{R}^+\right]}\,; \qquad x_2 > x_1\,,
\end{align}
in terms of the Wronskian
\begin{equation}
W \left[\tilde{R}^\infty, \tilde{R}^+\right] =   \tilde{R}^\infty \partial_x \tilde{R}^+ - \tilde{R}^+ \partial_x \tilde{R}^\infty\,,
\end{equation}
which depends on the frequency~$\omega$ but not on the radial coordinate~$x$.
At spatial infinity, the ingoing solution behaves as
$\tilde{R}^+ (x) \sim A_\text{\tiny in} (\omega) {\rm e}^{-{\rm i} \omega x} + A_\text{\tiny out} (\omega) {\rm e}^{{\rm i} \omega x}$, in terms of the complex-valued incidence and reflection coefficients~$A_\text{\tiny in,out}$, such that~$W = - 2 {\rm i} \omega A_\text{\tiny in} (\omega)$~\cite{Leaver:1986gd}. 

In order to understand the behavior of the retarded Green’s function in different time intervals, one can bend the integration contour in Eq.~\eqref{Greentransform} into the lower half of the complex~$\omega$ plane. Three different contributions arise: 

\begin{itemize} 

\item First, there is the high-frequency contribution, which consists of radiation that propagates directly from the source to the observer, and is associated with the large-frequency arc that closes the contour in the complex~$\omega$ plane. This contribution reduces to the free-space Green's function in the limit of vanishing BH mass~\cite{Leaver:1986gd}. 

\item The second contribution comes from quasinormal modes (QNMs). These are singularities in the lower half complex plane (as expected from a retarded Green's function), which occur at frequencies where the Wronskian vanishes, {\it i.e.}, where the two solutions become linearly dependent. They describe the interaction of the signal from the source with the curved BH geometry, and the fraction of the ingoing initial perturbation that is eventually reflected from the curvature potential near the event horizon back to spatial infinity~\cite{Leaver:1986gd, Berti:2009kk, Hui:2019aox}.

\item Lastly, there is a contribution describing the late-time behavior of the Green's function, where the dynamics starts to be affected by radiation reaction effects~\cite{Detweiler:2005kq, Poisson:2011nh, Barack:2018yvs}. Among them, tail effects 
depend on the asymptotic behavior of the effective potential~\cite{Ching:1995tj} and
account for the wave scattering off the weak Coulomb potential near infinity~\cite{Leaver:1986gd}. In other words, they account for the propagation of waves on the light-cones of the BH geometry in linear theory, rather than those of Minkowski spacetime. For example, for asymptotically flat spacetimes, the power-law decrease of the potential at spatial infinity gives rise to late-time tails of the retarded Green's function~\cite{PhysRevD.5.2419}, which are usually associated with the existence of a branch cut along the negative imaginary axis in complex frequency space.

\end{itemize}

\begin{figure}[t!]
	\centering
 	\includegraphics[width=0.48\textwidth]{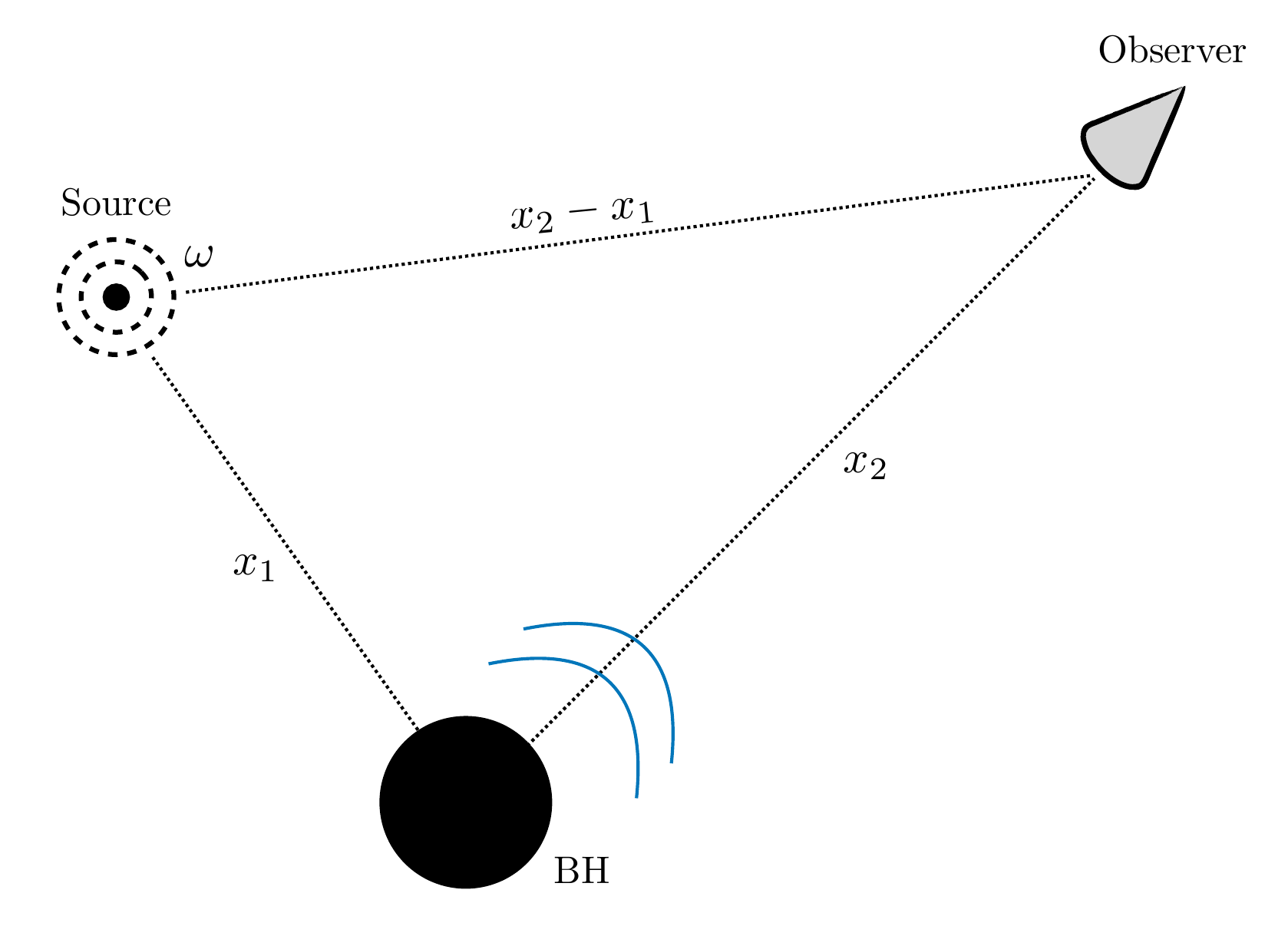}
	\caption{\it Pictorial representation of a BH response to
initial data of compact support. The latter is described by a source with characteristic frequency~$\omega$, located at a distance~$x_1$ from the BH. The observer, located at a distance~$x_2 \gtrsim x_1$, measures the BH response. The signal directly transmitted from the source to the observer arrives roughly at~$t_2 = x_2 - x_1$, while the response arrives at~$t_2= x_2 + x_1$, which is dominated by QNMs. At times~$t_2 \gg x_2 + x_1$, after the exponential decay due to the QNM ringing, the field will decay as a power law~\cite{Leaver:1986gd}.
}
	\label{Green}
\end{figure}

Of these three contributions, the first does not contain any information about the BH, while the other two encode the response of the object to the external source.
This is transparent in the asymptotic approximation, in which it is assumed that the spacetime is essentially flat around the observer and the source, both located far away from the BH ($x_2, x_1 \gg x_+$, but keeping the hierarchy~$x_2 \gtrsim x_1$). In this regime, the Green's function reduces to~\cite{Andersson:1996cm}
\begin{equation}\label{eq:AsymptoticGreen}
G(x_1,x_2;\omega) \simeq - \frac{1}{2 {\rm i} \omega} \llp {\rm e}^{{\rm i} \omega (x_2-x_1)} + \frac{A_\text{\tiny out}}{A_\text{\tiny in} } {\rm e}^{{\rm i} \omega (x_2+x_1)} \rrp\,.
\end{equation}
The first term highlights the large frequency regime, which does not carry any information about the BH, and depends only on the difference~$x_2 - x_1$, as predicted when the translation symmetry is restored at large distances. The second term instead captures the BH response, with QNMs given by the zeros of~$A_\text{\tiny in} (\omega)$. This is schematically shown in Fig.~\ref{Green} for the scales of interest. 

\section{Tidal response and Green's functions in Newtonian gravity}
\label{sec: TLN}
\noindent
Let us now review the basic formalism to extract the TLNs in {\it Newtonian gravity}. This will highlight the connection between tidal deformations and the retarded Green's function.

Consider a spherically symmetric spinning body of mass~$M$, assumed to be at the origin of a Cartesian coordinate frame, which is subjected to an external tidal gravitational field~$U_\text{\tiny ext}$ applied adiabatically. Within spherical symmetry, the gravitational field can be expanded in multipole moments as
\be\label{eq:UMultipole}
U_\text{\tiny ext} =-\sum_{\ell = 2}^\infty \frac{(\ell-2)!}{\ell!}r^\ell \mathcal{E}_{L}  n^L\,,
\ee
in terms of its distance~$r$ from the body, the normal to the sphere~$n^i \equiv x^i/r$, and the symmetric trace-free multipole moments~$\mathcal{E}_{L}$. As usual,
the multi-index~$L$ stands for~$\ell$ spatial indices,~$\mathcal{E}_L =\mathcal{E}_{(i_1\cdots i_\ell)}$, and~$n^L = n^{i_1}\cdots n^{i_\ell}$.

The presence of this external tidal force induces a deformation in the body, which gives rise to internal multipole moments as (setting~$G=1$ for simplicity)
\be
I_{L} = \int {\rm d}^3x~\delta\rho(\vec x) r^\ell  n^L \,,
\ee
where~$\delta\rho$ is the body's mass density perturbation.

Expanding the external source~$\mathcal{E}_{L}$ and induced response~$I_{L}$ in spherical harmonics~$Y_{\ell m}$,
\begin{eqnarray}
\nonumber
\mathcal{E}_{\ell m} & \equiv & \mathcal{E}_{L}\int {\rm d} \Omega~n^L Y^{*}_{\ell m}(\theta,\varphi)\,; \\
I_{\ell m} & \equiv & I_{L}\int {\rm d}\Omega~n^L Y^{*}_{\ell m}(\theta,\varphi)\,,
\end{eqnarray}
where~${\rm d} \Omega \equiv \sin\theta\,{\rm d}\theta\, {\rm d}\varphi$,
the total potential of the system becomes 
\be 
\label{potential}
U_\text{\tiny tot}=
-\frac{M}{r}
-\sum_{\ell, m} Y_{\ell m} 
\left[\frac{(\ell-2)!}{\ell !} \mathcal{E}_{\ell m}  r^\ell -
\frac{(2\ell-1)!!}{\ell !} 
\frac{I_{\ell m}}{r^{\ell+1}}
 \right]\,.
\ee
Under the assumption of adiabatic and weak external tidal forces, {\it linear response theory} dictates that the response multipoles are proportional to the perturbing multipole moments as
\be
\label{IkENew}
I_{\ell m}\left(\omega\right)
=-\frac{\left(\ell-2\right)!}{(2\ell-1)!!}k_{\ell m}(\omega) r_+^{2\ell+1} \mathcal{E}_{\ell m}\left(\omega\right)\,,
\ee
in terms of the perturbation frequency~$\omega$ and object size~$r_+$.
The dimensionless coefficients~$k_{\ell m}$ describe the tidal response and can be expanded as
\be
\label{eq:klm}
	k_{\ell m} \simeq \kappa_{\ell m} + {\rm i}\nu_{\ell m}\left(\omega - m \Omega\right) + \dots\,,
\ee
where~$m$ is the azimuthal harmonic number, and~$\Omega$ the body's angular velocity. The first real term describes the conservative and static response, with the coefficients~$\kappa_{\ell m}$ called TLNs, while the imaginary contribution~${\rm i}\nu_{\ell m}$ describes dissipative effects. 

We can now show how the static TLN appears inside the Green's function of the Newtonian problem. The Green's function satisfies the equation
\begin{equation}
\label{GFhomo}
\nabla^2_{x_2} G (\vec x_2,\vec x_1) = 4 \pi \delta^{(3)} (\vec x_2- \vec x_1)\,,
\end{equation}
where~$\vec x_1$ represents the position of the source, and~$\vec x_2$ that of the observer. The most general solution reads
\begin{equation}
G (\vec x_2,\vec x_1) = \frac{1}{|\vec x_2 - \vec x_1|} + G_\text{\tiny hom} (\vec x_2,\vec x_1)\,,
\end{equation}
where~$ G_\text{\tiny hom}$ is a solution of the homogeneous problem, to be fixed according to the boundary conditions of the physical problem at hand. Focusing on~$r_2 < r_1$, we can perform a spherical harmonic decomposition to obtain
\begin{align}
G (\vec x_2,\vec x_1) = & \sum_{\ell, m} \frac{4 \pi}{2\ell+1} \lp \frac{r_2^\ell}{r_1^{\ell+1}} + \frac{(2\ell-1)!!}{\ell !}  \frac{I_{\ell m} (r_1)}{r_2^{\ell+1}} \rp \nonumber \\
& ~~ \times Y_{\ell m} (\theta_2,\varphi_2)  Y_{\ell m}(\theta_1,\varphi_1)\,.
\end{align}
The first term is the expansion of~$\frac{1}{|\vec x_2 - \vec x_1|}$, while the second term comes from the homogeneous solution. Introducing the strength of the external source~$\mathcal{E}_{\ell m}(x_1) \propto 1/r_1^{\ell+1}$, the above can be rewritten as
\begin{align}
G (\vec x_2,\vec x_1) = & - \sum_{\ell, m} \lp \frac{(\ell-2)!}{\ell !} \mathcal{E}_{\ell m}(r_1) r_2^\ell - \frac{(2\ell-1)!!}{\ell !}  \right. \nonumber \\
& \left. \times \frac{I_{\ell m} (r_1)}{r_2^{\ell+1}} \rp  Y_{\ell m} (\theta_2,\varphi_2)  Y_{\ell m}(\theta_1,\varphi_1)\,.
\end{align}
Assuming linear response theory, one can relate the induced response to the external source~$I_{\ell m} (r_1) \propto - k_{\ell m} r_+^{2\ell +1}  \mathcal{E}_{\ell m} (r_1)$, as shown in Eq.~\eqref{IkENew}. We stress that this proportionality condition can alternatively be obtained by matching the inner and outer solutions of the perturbative problem at the boundary surface of the linearly perturbed object~\cite{PoissonWill}. The Green's function then reads
\begin{align}
\label{GF1stNew}
G (\vec x_2,\vec x_1) = & - \sum_{\ell, m}  \frac{(\ell-2)!}{\ell !} \mathcal{E}_{\ell m}(r_1) r_2^\ell \llp 1 + k_{\ell m} \lp \frac{r_+}{r_2} \rp^{2\ell+1} \rrp  \nonumber \\
& ~~ \times Y_{\ell m} (\theta_2,\varphi_2)  Y_{\ell m}(\theta_1,\varphi_1)\,.
\end{align}
This shows the direct connection between TLNs and Green's functions in the Newtonian theory.

\section{Tidal Love numbers in General Relativity}
\label{sec: TLN-GR}

\subsection{Static TLNs and effective field theory}
\label{sec: TLN-EFT}
\noindent
The above definition of TLNs within the Newtonian regime represents the nonrelativistic approximation to the full general relativity theory, in which one studies massless test fields propagating on a background geometry. 

To provide an explicit example, we focus on the propagation of scalar perturbations on a Kerr metric, which describes a rotating  BH of mass~$M$ and spin~$J$. In spherical coordinates~$(t,r,\theta,\varphi)$, the metric is given by
\begin{equation}
\d s^2 = - \frac{\Delta(r)}{r^2} \d t^2 +  \frac{r^2}{\Delta(r)} \d r^2  + r^2 \lp \d \theta^2 + \sin^2\theta  \d \varphi^2 \rp\,,
\end{equation}
where~$\Delta(r)$ is defined as
\begin{equation}
\Delta(r) = (r-r_+)(r-r_-)\,, 
\label{Delta def}
\end{equation}
in terms of the outer and Cauchy horizons. These are respectively given by
\begin{equation}
r_\pm = M \pm \sqrt{M^2 -\chi^2}\,,
\end{equation}
where we have introduced the spin parameter~$\chi = J/M$. 
Focusing on a single frequency mode~$\omega$, we decompose the scalar field~$\Phi$ as
\begin{equation}
\Phi(t,r,\theta,\varphi) = \sum_{\ell,m} R_{\ell m} (r) {\rm e}^{-{\rm i} \omega t} {\rm e}^{{\rm i} m \varphi} S_{\ell m}(\theta)\,,
\end{equation}
to obtain a system of equations for the angular and radial functions~\cite{Berti:2005eb}. 
The equation for the angular functions is
\begin{align}
\label{EQRN}
&\frac{1}{\sin \theta} \frac{\d}{\d \theta} \lp \sin \theta \frac{\d S_{\ell m}(\theta)}{\d \theta} \rp  \nonumber \\
& ~~~~~~~~ + \llp \lambda_{\ell m} +  \omega^2 \chi^2 \cos^2 \theta - \frac{m^2}{\sin^2\theta} \rrp S_{\ell m}(\theta) = 0\,,
\end{align}
in terms of separation constants~$\lambda_{\ell m}$. The solutions are spheroidal harmonics, which for~$\omega \chi\ll 1$ reduce to spherical harmonics~\cite{Goldberg:1966uu}. Making this assumption in the rest of the paper, the separation constant reduces to~$\lambda_{\ell m} = \ell (\ell+1) + 2 \chi m \omega + \mathcal{O}(\omega^2 \chi^2)$~\cite{Maldacena:1997ih}.
The equation for the radial functions is
\begin{align}
& \frac{\d}{\d r} \lp \Delta(r) \frac{\d R_{\ell m}(r)}{\d r} \rp \nonumber \\
& ~~~~~+  \left[\frac{\big(\omega \left(r^2+\chi^2\right)- m\chi\big)^2}{\Delta(r)} - \lambda_{\ell m}\right] R_{\ell m}(r) = 0\,.
\label{R eqn}
\end{align}
In terms of the rescaled function
\begin{equation}
\tilde R_{\ell m} (r) \equiv \sqrt{r^2 + \chi^2} R_{\ell m} (r)\,,
\end{equation}
Eq.~\eqref{R eqn} reduces to a Schrödinger-like equation of the form in Eq.~\eqref{Schrodinger}, namely
\begin{equation}
\frac{\d^2 \tilde{R}_{\ell m}}{\d x^2}  + \Big(\omega^2 - V_{\ell m}(x)\Big) \tilde{R}_{\ell m} = 0\,,
\label{Schrod like}
\end{equation}
where the tortoise coordinate~$x$ is defined by~$\frac{{\rm d} r}{{\rm d} x} = \frac{\Delta(r)}{r^2 + \chi^2}$.
The effective potential~$V_{\ell m}$ is given in the original~$r$ coordinate by~\cite{Kokkotas:2010zd}
\begin{align}
V_{\ell m}(r) & = \frac{1}{2r} \frac{\d }{\d r} \left(\frac{r^2 \Delta^2(r)}{(r^2 + \chi^2)^3}\right) + \omega^2 \nonumber \\
& - \frac{\Delta(r)}{(r^2 + \chi^2)^2} \left[\frac{\big(\omega \left(r^2+\chi^2\right)- m\chi\big)^2}{\Delta(r)} - \lambda_{\ell m}\right] \,.
\label{Vlm}
\end{align}
This potential has 4 singular points, located at the curvature singularity~$r = 0$, at the Cauchy and outer horizons~$r_\pm$, and at spatial infinity~$r = \infty$. At finite frequency, spatial infinity is an irregular singular point, and the problem admits confluent Heun-like solutions~\cite{Fiziev:2005ki,Bezerra:2013iha,Vieira:2014waa}.

In the static~($\omega = 0$) limit, the ingoing boundary condition implies regularity of the solution of the wave equation at the BH outer horizon~$r_+$. Near spatial infinity~($r \to \infty$), it is easy to see that Eq.~\eqref{Schrod like}, expressed in terms of the original variable~$R_{\ell m}(r)$, reduces to
\begin{equation}
\frac{\d^2 R_{\ell m}(r)}{\d r^2} + \frac{2}{r} \frac{\d R_{\ell m}(r)}{\d r} - \frac{\ell (\ell + 1)}{r^2} R_{\ell m}(r) = 0\,.
\end{equation}
Therefore the solution near spatial infinity takes the form
\be
\label{scalarTLN}
R_{\ell m}(r) \simeq C r^{\ell}\left[1+\cdots+ \kappa_\ell \left(\frac{r_+}{r}\right)^{2\ell+1}+\cdots\right]\,,
\ee
where the ellipses denote subleading powers of~$r_+/r$.
Following the Newtonian matching procedure, one can make an analogy with the nonrelativistic gravitational potential of Eq.~\eqref{potential} to recognize the external tidal field in the growing  term~$\sim r^{\ell}$, and the BH response in the decaying one~$\sim r^{-\ell-1}$. The coefficients~$\kappa_\ell$ denote the static TLNs defined in Eq.~\eqref{eq:klm}. These are found to vanish identically for Kerr geometries~\cite{Hui:2020xxx}, as we will review in Sec.~\ref{sec: effectivegeometry}.\footnote{The overlap between the source series and the response contribution in the physical case~$\ell \in \mathbb{N}$~\cite{Kol:2011vg,LeTiec:2020spy,Charalambous:2021mea} requires us to perform an analytic continuation to the unphysical region~$\ell \in \mathbb{R}$~\cite{LeTiec:2020spy}, where the source and response series do not overlap.}

\bigskip
Tidal effects  can also be described in the point-particle {\it effective field theory} (EFT) formalism~\cite{Goldberger:2005cd, Chakrabarti:2013lua, Porto:2016pyg, Goldberger:2020wbx, Goldberger:2020fot, Creci:2021rkz}. This framework is based on the realization that a BH behaves as a point particle at large distances~$r \gg r_+$, and corrections due to its internal structure are encoded in higher-derivative operators in the effective theory. 

To illustrate the structure of the coupling between the multipole moment~$I_L$ and external tidal fields~$\mathcal{E}_L$ in this context, we consider a scalar field on a flat background coupled to an external source~\cite{Ross:2012fc}
\be
S = \int {\rm d}^4 x \lp \frac12 \, \partial^\mu \phi \, \partial_\mu \phi + J \, \phi  \rp  \,.
\ee
The source~$J(t,\vec{x})$ represents the skeletonized BH, and we consider the case where the spatial gradient of~$\phi$ is small relative to the size of the source. We can Taylor expand the scalar field in the interaction Lagrangian around a point inside the source, chosen as the origin ($\vec{x} = 0$):
\begin{align}
    S_\text{\tiny int} &= \int {\rm d}^4 x \, J(t,\vec{x}) \, \sum_{n=0}^\infty \, \frac{x^{k_1} \dots x^{k_n}}{n !} \big( \partial_{k_1} \dots \partial_{k_n} \phi\big)\big\vert_{\vec{x}=0} \nonumber \\
    &= \int {\rm d} t \, \sum_{n=0}^\infty \, \frac{1}{n!} \, M^{k_1 \dots k_n}\, \big( \partial_{k_1} \dots \partial_{k_n} \phi\big)\big\vert_{\vec{x}=0}\,,
\end{align}
where the symmetric (not yet traceless) source moments are defined as $M^{k_1 \dots k_n} \equiv \int {\rm d}^3 x \, J(t, \vec{x}) \, x^{k_1} \dots x^{k_n}$. Following~\cite{Ross:2012fc}, we can rearrange the tensor $M^{k_1 \dots k_n}$ in a symmetric trace-free (STF) fashion
\be
\label{eq:ActionInt_MultipoleExpansion}
S_\text{\tiny int} = \int {\rm d} t \int {\rm d}^3 y \, \sum_{\ell = 0}^\infty \, \frac{\delta^3(\vec{y})}{\ell!} I_L(t) \, \partial_L \phi(t, \vec{y})\,, 
\ee
where~$L$ is the STF index notation adopted earlier, we have introduced the factor of~$\delta^3(\vec{y})$ resulting from the evaluation of the scalar field derivative at the origin, and where the multipole moment is defined as 
\be 
I_L(t) = \sum_{p = 0}^\infty \frac{(2 \ell + 1 )!!}{(2 p)!! (2 \ell + 2 p + 1 )!!} \int {\rm d}^3 x \, r^{2 p} x^L  \partial_t^{2 p} J(t, \vec{x}) \,.
\ee 
The form of the action in Eq.~\eqref{eq:ActionInt_MultipoleExpansion} highlights the structure of the tidal interaction between the external field~$\mathcal{E}_L \propto \partial_{L} \phi$, and the source with multipole moments~$I_L$. 

The scalar field equation,
\be
\Box\phi(t,\vec{x}) = \sum_{\ell = 0}^\infty \frac{(-1)^\ell}{\ell!} I_L(t) \partial_L \delta^3(\vec{x})\,,
\ee
can be solved in terms of the retarded Green's function, 
\be 
G (t_2, \vec{x}_2; t_1, \vec{x}_1) = \frac{- {\rm i} \theta (t_2 - t_1)}{4 \pi |\vec{x}_2 - \vec{x}_1|} \delta \big( t_2  - t_1 - |\vec{x}_2 - \vec{x}_1| \big)\,.
\ee 
Explicitly, 
\begin{align} 
\phi (t, \vec{x}) & = - \sum^\infty_{\ell = 0} \,\frac{1}{\ell!}\Bigg\{(\ell-2)!\,\mathcal{E}_L x^L \nonumber \\
& - {\rm i}  \int {\rm d}^4 y \, \delta^3(\vec{y}) \, I_L(y^0) \, \frac{\partial}{\partial y^L}\, G (t, \vec{x}; y^0, \vec{y})\Bigg\}\,,
\end{align}
where the homogeneous solution (first line) describes the external tidal field, as before.
Performing the derivatives on the Green's function\footnote{A useful relation is
\be
\frac{\partial \delta \lp t  - y^0 - |\vec{x} - \vec{y}| \rp}{ \partial y^k}  =  \frac{(x^k - y^k)}{|\vec{x} - \vec{y}|} \lp - \frac{\partial \delta \lp t  - y^0 - |\vec{x} - \vec{y}| \rp}{\partial y^0}\rp\,.
\ee
}
leads to
\begin{align}
\label{phiEFT}
\phi (t, \vec{x}) & = - \sum^\infty_{\ell = 0} \, \frac{x^L}{\ell!}\Bigg\{ (\ell-2)! \, \mathcal{E}_L  \nonumber \\
& -   \, \frac{(2\ell-1)!!}{r^{2 \ell + 1} }\, \sum^\ell_{n=0} {\cal C}_{\ell n} \, r^n  \frac{\partial^n I_L(y^0)}{\partial {y^0}^n}\Bigg|_{y^0 = t - r}\Bigg\}\,,
\end{align}
where the explicit form of the combinatorial coefficients~$C_{\ell n}$, which depend on $\ell$ and~$n$, will not be needed.

\begin{figure*}[t!]
	\centering
 	\includegraphics[width=1\textwidth]{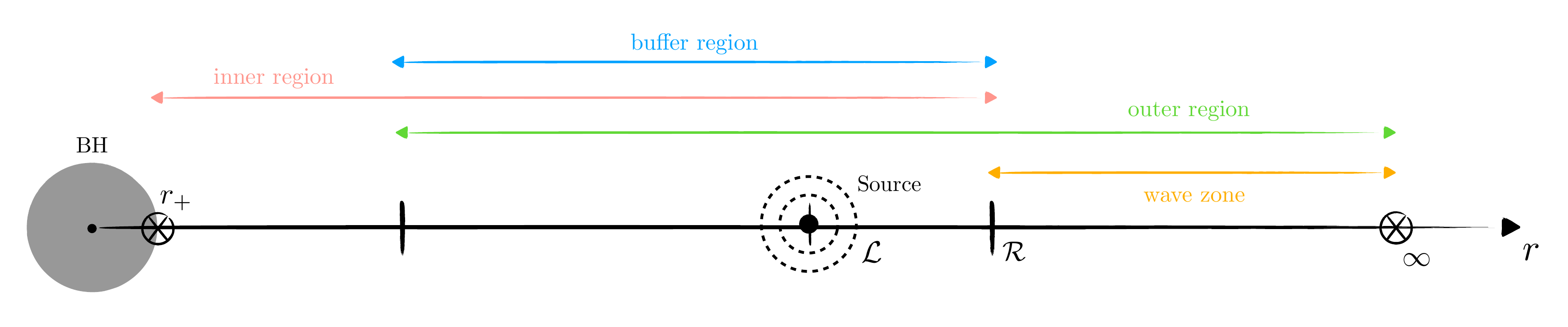}
	\caption{\it Sketch of the physical scales involved in the problem. A source of characteristic frequency~$\omega \sim 1/\mathcal{R}$, located at a distance~$\mathcal{L}$, tidally deforms a BH of radius~$r_+$. The rose arrow describes the inner region~($r_+ \lesssim r \lesssim \mathcal{R}$); the green arrow denotes the outer region~($r \gg r_+$), which includes spatial infinity; the blue arrow indicates the overlap between them in the buffer region; and the orange line identifies the wave zone~($r \gtrsim \mathcal{R}$).}
	\label{scales}
\end{figure*}

At this point, we can adopt linear response theory to show that the external tide induces a nonvanishing expectation value for the moment operator~$I_L$ as
\be
\label{vev}
\langle I_L(t)\rangle = \int \d t' G^I_{LL'}(t,t')  \mathcal{E}_{L'}(x(t'))\,,
\ee
where the brackets~$\langle ...\rangle$ denote an ensemble-averaging over short-scale modes (setting to zero the external tidal field). In the above we have introduced a retarded Green's function for the moment field $I_L$ as
\be 
G^I_{L L'}(t,t') = - {\rm i} \langle [I_L(t),I_{L'}(t') ]\rangle \, \theta(t-t')\,.
\ee
This expression is the gravitational analog of the Kubo formula~\cite{denittis2016linear}.
In the frequency domain, the analyticity of the Green's function around~$\omega=0$ (compatible with the requirement~$\omega r_+ \ll 1$ for the validity of the EFT), 
allows us to write~\cite{Porto:2016pyg, Goldberger:2020fot}
\be 
\label{Gana}
G^I_{L L'}(\omega)=
\sum_{n=0}^\infty
\delta_{L L'}
\omega^{2n} 
\left(
\lambda_{2n}^{(L)}
+ {\rm i} \, \lambda_{2n+1}^{(L)}
\omega
\right)\,,
\ee
where the first terms are symmetric under time-reversal
symmetry and describe the conservative dynamics, while the second terms describe dissipative effects. Let us stress that Eq.~\eqref{Gana} describes only the {\it instantaneous} response, {\it i.e.}, when the internal dynamics is fast compared to the timescale of the tidal perturbation. This statement is equivalent to assuming exponentially decaying correlation functions at late times; that is, in the absence of long time tails~\cite{Goldberger:2005cd,Goldberger:2020wbx,Goldberger:2020fot}. In the full dynamical process, the Green's function will also receive delayed contributions from tail effects, which we will discuss later in the paper.

Equation~\eqref{vev} can then be used to simplify the scalar field solution of Eq.~\eqref{phiEFT}. For instance, consider the static case $G^I_{L L'} (t,t') = \lambda_0^{(L)} \delta (t-t') \delta_{LL'}$, and the first term $n = 0$ in the series. In this case, we recover the usual factorization into growing and decaying behaviors as
\begin{align}
\label{eq:EFTScalarField}
\phi (t, \vec{x}) = - \sum^\infty_{\ell = 0} \, \frac{(\ell-2)!}{\ell!} \mathcal{E}_L x^L \left[ 1 - \frac{1}{4\pi \left(\ell-2\right)!} \frac{\lambda_0^{(L)}}{r^{2 \ell + 1}} \right]\,,
\end{align}
which shows the appearance of the Wilson coefficients $\lambda_0^{(L)}$ of the tidal operators in the scalar field solution. As shown in Ref.~\cite{Hui:2020xxx}, the direct comparison between the EFT result and the one from general relativity, shown in Eq.~\eqref{scalarTLN}, allows us to relate the static TLN $\kappa_{\ell}$ to the Wilson coefficient of the relevant operator.

 Before closing this section, it is interesting to outline the connection between the TLNs and the Green's function from the EFT perspective. Following the steps discussed in the Newtonian case, we can immediately realise that the Green function of the static EFT problem takes the form
\begin{align}
\label{GFEFT}
G (\vec{x}_2,\vec{x}_1) &= - \sum^\infty_{\ell = 0} \, \frac{(\ell-2)!}{\ell!} \mathcal{E}_L (x_1) x_2^L  \nonumber \\
& ~~~~~\times \left[1 - \frac{1}{4\pi \left(\ell-2\right)!} \frac{\lambda_0^{(L)}}{r_2^{2 \ell + 1}} \right]\,,
\end{align}
where we have identified an external source $\mathcal{E}_L (x_1)$ from the STF decomposition of the standard $1/|\vec{x}_2 - \vec{x}_1|$ contribution, differently than in Eq.~\eqref{eq:EFTScalarField},  where ${\cal E}_L \propto \partial_L \phi$ depends on the field configuration.
 As a final comment, one can also recognize that the bulk Green's function $G (\vec{x}_2,\vec{x}_1)$ is proportional to the Green's function~$G^I_{L L'}$ of the multipole moments, introduced in Eq.~\eqref{Gana} in the context of response theory.
Such a connection confirms that the bulk Green's function ultimately contains all the information concerning the tidal response of the system.

\subsection{Dynamical response and asymptotic matching}
\label{sec: effectivegeometry}
\noindent
As discussed in the previous section, the response to tidal sources is captured by a set of coefficients, which describe both the static and dynamical response to external perturbations. The latter is based on frequency dependent perturbers~\cite{Perry:2023wmm}, and usually appears as the physically relevant scenario in the context of binary systems during the inspiral phase. Finding the dynamical response has, however, been shown to suffer from a major drawback due to the difficulty of solving the perturbation equations at finite frequency, which has been achieved only in certain limiting regimes. 

The standard approach to treat time-dependent tidal excitations of spherically symmetric BHs invokes the technique of matched asymptotic expansions~\cite{Poisson:2020vap, HegadeKR:2024agt}, based on the separation of scales schematically shown in Fig.~\ref{scales}. This formalism relies on solving the perturbation equations in three distinct spacetime regions, and matching the solutions inside regions of common overlap. Before describing these regions, let us stress that this approach is also based on a perturbative series in the parameter~$\omega M \ll 1$, which is equivalent to an expansion in the strength of the external tidal source.

The different regions are defined in terms of three physical scales:
\begin{itemize}

\item[--] The radius of curvature,~$\mathcal{R}$, of the external spacetime in which the body is inserted.

\item[--] The scale~$\mathcal{L}$ of spatial inhomogeneity of the curvature.

\item[--] The scale~$\mathcal{T}$ of temporal variation.

\end{itemize}
In the context of tidal perturbations from an external source, both~$\mathcal{R}$ and~$\mathcal{T}$ are $\sim \omega^{-1}$, while~$\mathcal{L}$ is determined by the distance to the source. In the following we will assume~$\mathcal{L} \ll \omega^{-1}$,  such that corrections from the gradient expansion of the metric dominate over those from the curvature and time derivative expansions~\cite{Poisson:2020vap, HegadeKR:2024agt}.

The different regions of interest are (see Fig.~\ref{scales}):

\begin{itemize} 

\item The {\it inner region}, $r_+ \lesssim r \lesssim \mathcal{R}$, in which the metric of the system is approximately given by the object metric with a small perturbation caused by the external source that is perturbing the BH. The inner region, beyond which the multipole expansion of the tidal field and the EFT approach break down, overlaps with the 
{\it outer region}. 

\item The {\it outer region}, at distances far from the BH,~$r \gg r_+$ (thus including spatial infinity), where the metric is approximately given by the background geometry of the external source, with a small perturbation caused by the BH. 

\item The {\it buffer region}, where the inner and outer regions overlap, corresponds to~$\left(\frac{r_+}{\mathcal{L}}\right)^{1/4} \mathcal{L}\lesssim r \lesssim \mathcal{R}$~\cite{Detweiler:2005kq}. The existence of a buffer region allows us to expand the inner region solution~$\tilde{R}^+$ at large radii to extract the growing and decaying branches, and read off the tidal coefficients.

\end{itemize}

Let us briefly review the computation of the dynamical response for Kerr BHs based on the effective near-zone geometry~\cite{Maldacena:1997ih, Hui:2022vbh}.
In the equation of motion~\eqref{Schrod like}, one keeps the~$\omega$-dependent terms that are proportional to~$1/\Delta$, to preserve the nature of the singularity at~$r_+$, while making the replacement~$r \to r_+$ in the corresponding numerators~\cite{Maldacena:1997ih}. For instance,~$\frac{\omega^2 r^4}{\Delta} \to \frac{\omega^2 r_+^4}{\Delta}$.\footnote{Note that a generic near-zone approximation consists of~$\frac{\omega^2 r^4}{\Delta} \to \frac{\omega^2 r_+^4 g(r)}{\Delta}$, with~$g(r \to r_+) = 1$ to preserve the dynamics near~$r_+$, and~$\lim_{r \to \infty} \frac{g(r)}{r^2} = 0$ to maintain the index structure~$r^\ell$~\cite{Charalambous:2024gpf}.} This ensures small corrections at the horizon and away from it in the low frequency regime ~$\omega \chi \leq \omega r_+ < \omega r  \ll 1$, which easily fits within the region~$r/\mathcal{R} \ll 1$. Evidently, the error associated with this approximation is of order~$\frac{\omega^2(r^4-r_+^4)}{\Delta}$, which can be evaluated at the onset of the buffer region:
\begin{equation}
\frac{\omega^2(r^4-r_+^4)}{\Delta}\bigg|_{\left(\frac{r_+}{\mathcal{L}}\right)^{1/4}\mathcal{L}} \simeq \mathcal{L}^2 \omega^2 \sqrt{\frac{r_+}{\mathcal{L}}} \lesssim \sqrt{\frac{r_+}{\mathcal{L}}} \ll 1\,,
\end{equation}
where in the second step we have used that~$r_+ \ll \mathcal{L}$ and  bounded~$\mathcal{L} \lesssim \mathcal{R} \sim 1/\omega$. 

A key implication of this replacement is to change the nature of the irregular singularity at spatial infinity, which now becomes regular. Defining~$z = \frac{r - r_+}{r-r_-}$, one can show that the wave equation~\eqref{Schrod like} reduces to
\begin{align}
\label{hypergeoS}
& z(1-z) R_{\ell m}''(z) + (1-z) R_{\ell m}'(z)  \nonumber \\
&~~~~ + \left[ \xi^2\left(\frac{1}{z} - 1\right) - \frac{\ell(\ell+1)}{1-z} \right] R_{\ell m}(z) = 0\,, 
\end{align}
with~$' = \frac{{\rm d}}{{\rm d}z}$, and~$\xi \equiv \frac{\omega (r_+^2+\chi^2)-m\chi}{r_+ - r _ -}$.  
This is a {\it hypergeometric} equation, with singular points at the inner ($z \to \infty, \, r \to r_-$) and outer ($z \to  0, \, r \to r_+$) horizons, and at spatial infinity $(z \to 1, \, r \to \infty$), with two independent solutions in the neighborhood of each singular point. 

Imposing the boundary condition of ingoing waves near the outer horizon, the relevant solution around~$z = 0$ reads~\cite{Kehagias:2022ndy,Perry:2023wmm}  
\begin{equation}
\label{solhorRN}
R_{\ell m}^+ (z) = C z^{-{\rm i} \xi} (1-z)^{-\ell} \, {}_2 F_1 (a,b;c;z)\,,
\end{equation}
where~${}_2 F_1$ is the hypergeometric function of the $2{\rm nd}$ kind,~$C$ is a constant, and
\be
a = - \ell \,;\quad b = - 2{\rm i} \xi -\ell \,; \quad  c = - 2{\rm i} \xi + 1\,.
\ee
This solution matches the leading term in the series obtained with the Mano-Suzuki-Takasugi (MST) method~\cite{Mano:1996gn,Mano:1996mf, Mano:1996vt}. In the MST approach, the inner region solution is given by an infinite series of hypergeometric functions, while the outer region solution is based on an infinite series of Coulomb wavefunctions~\cite{Mano:1996gn,Mano:1996mf, Mano:1996vt}. 

We can now determine the TLNs associated to the solution~\eqref{solhorRN}. Following the asymptotic matching prescription, we can expand the solution of the inner region  to large radii in the buffer region, in order to recognize the growing and decaying branches\footnote{Recognizing the growing and decaying branches in the field solution relies on the assumption of weak tidal perturbations, which is hypothesized within the asymptotic matching procedure. In the time dependent problem (see footnote~1 for the static case), going to higher orders in frequency  may result into a source-response mixing, complicating the extraction of the dynamical TLNs. Similarly, gauge issues, associated to the matching of the field solutions of the inner and outer regions~\cite{Poisson:2020vap, HegadeKR:2024agt}, could bring further ambiguities.}. In practice, this amounts to 
using the Kummer property~\cite{abramowitz+stegun}
\begin{align}
\label{Kummer}
 &  {}_2 F_1 (a,b;c;z) \nonumber \\
 & =  \frac{\Gamma (c) \Gamma (c-a-b)}{\Gamma(c-a) \Gamma (c-b)} {}_2 F_1 (a,b;a+b-c+1;1-z) \nonumber \\
& ~~+ \frac{\Gamma (c) \Gamma (a+b-c)}{\Gamma(a) \Gamma (b)} (1-z)^{c-a-b} \nonumber \\
& ~~~~\times \, {}_2 F_1 (c-a,c-b;c-a-b+1;1-z)\, .
\end{align}
This allows us to express one of the solutions of the hypergeometric problem around one singularity in terms of the solutions around another singularity. Using~\eqref{Kummer} in~\eqref{solhorRN}, and keeping the leading terms near~$z = 1$, we obtain
\begin{align}
\label{ABBTZ}
R_{\ell m}^+ (z) & \simeq C \Gamma (c) \left[ \frac{ \Gamma (c-a-b)}{\Gamma(c-a) \Gamma (c-b)} \, (1-z)^{-\ell} \right. \nonumber \\
& \left. ~~~~~~~~~~~~ + \frac{ \Gamma (a+b-c)}{\Gamma(a) \Gamma (b)} \, (1-z)^{\ell+ 1} \right] \,.
\end{align}
Since~$1-z\simeq \frac{r_+-r_-}{r}$, this solution is of the form~$R_{\ell m}^+(r) = A \, r^\ell + B \, r^{-\ell-1}$.  

Following the standard definition of the tidal response as the ratio between coefficients of the decaying (response) and growing (source) branches, one can read this off from~\eqref{ABBTZ} as:
\begin{align}
k_{\ell m} & = \frac{\Gamma(c-a) \Gamma (c-b) \Gamma (a+b-c)}{\Gamma(a) \Gamma (b) \Gamma (c-a-b)} \nonumber \\
&= \frac{\Gamma \big(\ell+1 - 2 {\rm i} \xi \big)  \Gamma (\ell+1) \Gamma (-2 \ell-1)}{\Gamma (-\ell)  \Gamma\big(-\ell - 2{\rm i} \xi\big) \Gamma (2 \ell+1)}\,.
\end{align}
It is instructive to expand this expression for small spin~$\chi$ and frequency~$\omega$, using the expression for~$\xi$ below Eq.~\eqref{hypergeoS}.
The result is~\cite{Chia:2020yla,Charalambous:2021mea,Kehagias:2022ndy}:
\begin{align}
k_{\ell m}  &  \simeq
\frac{ \Gamma \left(-\ell-\frac{1}{2}\right) \Gamma (\ell +1)}{4^{2 \ell+1}\Gamma \left(\ell+\frac{1}{2}\right) \Gamma (-\ell)\tan \pi \ell}\nonumber \\
& ~~~~\times \left[ \tan \pi  \ell +  
\frac{2 {\rm i} \pi  \left(r_+^2 \omega - m \chi\right) }{r_+-r_-}
 \right] \,.
\end{align}
For integer~$\ell$, the overall prefactor is finite, hence the first term (real part) vanishes. Matching to~\eqref{eq:klm}, this implies that the conservative response of Kerr BHs vanishes in the static limit,~$\kappa_{\ell m} = 0$. The second term (imaginary part), which encodes the dissipative part~$\nu_{\ell m}$, is nonvanishing and proportional to the frame-dragging term~$\omega r_+^2 - m \chi$~\cite{Chia:2020yla,Charalambous:2021mea,Kehagias:2022ndy}. 
Finally, the dynamical TLNs for Schwarzschild BHs are found to vanish  within this approximation~\cite{Charalambous:2021mea,Poisson:2020vap}, even though $\mathcal{O}(\omega^2 r_+^2)$ corrections to the approximated equation of motion could affect this result. In fact, in the context of scattering amplitudes, the dynamical TLNs were instead found to be nonvanishing and characterised by a logarithmic running~\cite{Saketh:2023bul}.

\section{Tidal response and tail effects in the buffer region}
\label{sec: TLN&tail}
\noindent
As shown in Sections~\ref{sec: TLN} and \ref{sec: TLN-EFT}, the connection between TLNs and retarded Green's functions seems immediate in Newtonian gravity or in effective field theories. One may therefore inquire if such link is manifest even in the context of general relativity. 

The method of matched asymptotic expansions reviewed above allows  to build up a multipolar expansion within the buffer region,  where one can identify the growing and decaying branches, and to extract the corresponding TLNs. 
As shown in Fig.~\ref{scales}, the buffer region also overlaps with the outer region, which spans distances far from the BH and carries information about spatial infinity, contrarily to the inner-region  where the associated singularity is removed following the effective geometry approach discussed in Sec.~\ref{sec: effectivegeometry}.

In light of the discussion in Sec.~\ref{sec: GF}, the construction of retarded Green's functions in general relativity demands to consider also the outgoing solution near spatial infinity. Such solution is determined in the outer region and takes into account additional effects, such as radiation reaction and tail effects, that come into play at large distances and are not present in the inner-region solution. These effects contaminate the direct extraction of the tidal response from retarded Green's functions. 

In the following, we will investigate the problematics in the extraction of TLNs and dissipative coefficients from retarded Green's functions in general relativity, associated to the presence of radiation effects in the buffer region.

\subsection{Radiation reaction and tail effects}
\label{sec: RR}
\noindent
Radiation reaction accounts for the contribution of radiative modes of the outer region to the dynamics in
the inner region. It can be accounted for in an EFT approach by integrating out the wave zone contributions.
Similarly to the extraction of the dynamical response, the theory of the gravitational self-force in curved spacetime is based on the formalism of matched asymptotic expansions and singular perturbation theory~\cite{Kevorkian1996MultipleSA}. In that context, one must glue the metric of the BH perturbed by the external universe (source) through which it is moving, with the metric of the external universe once perturbed by the BH moving through it~\cite{Detweiler:2005kq}.

The relevant solution to the wave equation is the retarded one, since it properly incorporates outgoing boundary conditions at spatial infinity (contrary to the advanced solution based on ingoing conditions). The retarded (as well as the advanced) field is singular on the BH worldline, for the obvious reason that the gravitational potential of a point particle diverges at its position. 

In this context it is therefore useful to split the retarded solution into {\it singular} and {\it regular} fields:

\begin{itemize}

\item The {\it singular} field has the same singularity structure as the retarded solution and, by design, is insensitive to the boundary conditions at spatial infinity. That is, it corresponds to a vanishing net radiation flux at infinity.

\item The {\it regular} field is a smooth vacuum solution, and contains the backscattered waves that arise from propagation within the light cones of the background spacetime.

\end{itemize}
This split can be implemented at the level of retarded/advanced Green's functions as follows:
\begin{align}
G_\text{\tiny sing} &= \frac{1}{2} \left(G_\text{\tiny ret} + G_\text{\tiny adv} \right) \,; \nonumber \\
G_\text{\tiny reg} & = \frac{1}{2} \left(G_\text{\tiny ret} - G_\text{\tiny adv} \right)\,,
\label{Gsingreg}
\end{align}
such that~$G_\text{\tiny ret}  = G_\text{\tiny sing} +  G_\text{\tiny reg}$.\footnote{In curved spacetime, the construction of the singular Green's function~$G_\text{\tiny sing}$ demands the addition of a specific solution to the homogeneous equation satisfying characteristic boundary conditions on the null cones emanating from the source location~\cite{Poisson:2011nh}.} 

Clearly, the BH motion is intimately tied to the boundary conditions at spatial infinity. For instance, if the waves are outgoing the particle loses energy to radiation, while if the waves are incoming the particle gains energy from the radiation. Therefore, it should be possible to remove~$G_\text{\tiny sing}$ (which as mentioned above is insensitive to boundary conditions) without affecting the BH motion. The subtraction leaves behind the regular two-point function, which produces a field that is regular on the worldline and governs the motion of the particle. The behavior of this field is clearly delineated, accurately embodying the boundary conditions for outgoing waves: as a result, the particle will dissipate energy through radiation~\cite{Poisson:2011nh}. In other words, the BH reacts to both the perturbing source, as well as the self-force generated by the emitted gravitational waves.

The singularity structure of the two solutions suggests that the tidal effects on the BH are expected to be included in the singular solution. This is also supported by the fact that TLNs are ultimately dictated by the ingoing boundary condition at the BH horizon within the inner region, similarly to the singular contribution in the retarded Green's function. 
The contribution from TLNs may, however, be overshadowed by the regular contributions in the retarded solution. A proper matching procedure has to be performed in order to disentangle these two contributions in the full solution within the buffer region.

To clearly illustrate this conjecture, let us investigate the presence of {\it tail effects} on BH backgrounds. These effects arise from the nonlinear interactions between the emitted gravitational waves and the curved background, leading to delayed contributions in the gravitational waveform. They were studied for various BH geometries in Refs.~\cite{Leaver:1986gd,Andersson:1996cm, Barack:1999ma,Burko:2002bt,Casals:2015nja,Casals:2016soq,Konoplya:2013rxa}.

The late-time tail can be deduced by studying the low-frequency contribution to the Green’s function which, as we show below, is characterized by a branch cut along the negative imaginary axis in complex-frequency space. For this purpose, we focus for simplicity on Schwarzschild BHs, such that~$r_- = 0$,~$r_+ = 2M = r_s$, and~$\Delta(r) = r^2 f(r)$, with~$f(r) = 1 - r/r_s$ as usual.
Defining~$\hat{R}_{\ell m} = \sqrt{f(r)} \tilde{R}_{\ell m}$, the radial equation~\eqref{Schrod like} becomes
\begin{equation}
\frac{\d^2\hat{R}_{\ell m}}{\d r^2} + \left[\omega^2 + \frac{r_s^2}{4 r^4} - \frac{\ell (\ell+1)f(r)}{r^2} \right] \frac{\hat{R}_{\ell m}}{f^2(r)} = 0\,.
\end{equation}
Since tail effects involve backscattering off the slightly curved spacetime far from the BH, one can focus on distances~$r\gg r_s$ (that is, in the outer region). In this region, the above equation simplifies to
\begin{align}
\label{EQinfinity}
\frac{\d^2\hat{R}_{\ell m}}{\d r^2} +  \llp \omega^2\left(1 + \frac{2r_s}{r}\right) - \frac{\ell (\ell+1)}{r^2} \rrp \hat{R}_{\ell m} = 0\,,
\end{align}
where we have kept the leading~$r_s/r$ correction. The form of the equation shows the presence of an irregular singularity at spatial infinity, which implies that the general solution expressed in terms of~$\tilde{R}_{\ell m}$ is given by irregular confluent hypergeometric functions~\cite{Andersson:1996cm}:
\begin{align}
\label{Sinfty}
\tilde{R}_{\ell m} (r) &= C_1 \left( \frac{r}{r_s} \right)^{\ell+1} {\rm e}^{-{\rm i} \omega r} M (\ell + 1 + {\rm i} \omega r_s, 2 \ell + 2;  2 {\rm i} \omega r) \nonumber \\
& + C_2 \left( \frac{r}{r_s} \right)^{\ell+1} {\rm e}^{{\rm i} \omega r} U (\ell + 1 - {\rm i} \omega r_s, 2 \ell + 2; - 2 {\rm i} \omega r)\,.
\end{align}
Thus, we recognize that the ingoing solution is proportional to the confluent hypergeometric~$M$, while the outgoing solution is proportional to~$U$.
That is,
\begin{align}
\tilde{R}_{\ell m}^+ (r) & = C_1 \left( \frac{r}{r_s} \right)^{\ell+1} {\rm e}^{-{\rm i} \omega r} M (\ell + 1 + {\rm i} \omega r_s, 2 \ell + 2;  2 {\rm i} \omega r) \,; \nonumber \\
\tilde{R}_{\ell m}^\infty (r) & = C_2 \left( \frac{r}{r_s} \right)^{\ell+1} {\rm e}^{{\rm i} \omega r} U (\ell + 1 - {\rm i} \omega r_s, 2 \ell + 2; - 2 {\rm i} \omega r)\,.
\end{align}
This is compatible with the solution obtained in the MST approach~\cite{Mano:1996gn, Mano:1996mf, Mano:1996vt}. In the limit~$\omega r_s \ll 1$, one easily recovers the known Bessel functions of asymptotically flat spacetimes.\footnote{In the context of a scattering scalar wave on a BH background, this solution can be used to relate the tidal response to the reflection amplitude~\cite{Ivanov:2022qqt, Charalambous:2024gpf}, showing that vanishing TLNs can be interpreted as reflectionless, total transmission modes~\cite{Hod:2013fea, Cook:2016fge, Cook:2016ngj}.}

One can construct the corresponding retarded Green's function, following the approach in Sec.~\ref{sec: GF} (see Eq.~\ref{eq:GreenFunction_Definition}), as\footnote{In the asymptotic region~$r \gg r_s$, we can replace the tortoise coordinate~$x$ with the radial coordinate~$r$.}
\begin{align}
    & G(r_1 < r_2; \omega) = \frac{(-{\rm i})^{2 \ell + 1} (\omega r_s)^{2(\ell+1)} \Gamma \lp \ell + 1 - {\rm i} \omega r_s\rp}{ \omega  \,  \Gamma \lp 2 \ell + 2\rp} \nonumber \\
   &~~~~~~~~~~ \times \tilde{R}_{\ell m}^+ (r_1)  \tilde{R}_{\ell m}^\infty (r_2)\,.
\end{align}
This can be expanded in the buffer region according to~$\omega r_s \ll \omega r_{1,2} \ll 1$ to obtain, at first order in~$\omega r_s$,
\begin{align}
\label{Greenouter}
  &   G(r_1 < r_2; \omega)  \simeq \frac{ \Gamma\left(-\ell-\frac{1}{2}\right)}{2 \omega (2{\rm i})^{2 \ell + 1}\Gamma(\ell+3/2)} \nonumber \\
    & ~~~ \times (\omega r_1)^{\ell+1} (\omega r_2)^{\ell+1} \llp 1 - \frac{ (-2\rm i)^{2\ell+1}}{(\omega r_2)^{2\ell+1}} \frac{\Gamma\left(\ell+\frac{1}{2}\right)}{\Gamma\left(-\ell-\frac{1}{2}\right)}\rrp \nonumber \\
    & ~~~ \times \bigg( 1 + {\rm i} \omega r_s \Big[ {\rm log}(- 2 {\rm i}\omega r_2) - \psi (\ell+1)\Big] \bigg)\,,
\end{align}
where~$\psi (z) = \Gamma'(z)/\Gamma(z)$ denotes the digamma function. 
At order~$(\omega r_s)^0$, one is left with the first two lines, where one recognizes a factorization of the Green's function into a source (growing terms) and a response (decaying terms), similar to the asymptotic expression shown in Eq.~\eqref{eq:AsymptoticGreen}. However, the nonzero BH mass brings an additional contribution to the radial dependence, proportional to a logarithmic term. This is in contrast with the instantaneous decomposition of the Green's function outlined within the EFT approach (see Eq.~\eqref{Gana}), which makes the extraction of the tidal response unclear.

The logarithmic term traces back to the confluent hypergeometric function~$U$. It implies the existence of a branch cut, starting at the branch point~$\omega = 0$ and running along the negative frequency imaginary axis.\footnote{Notice that in the strict static case, $\omega = 0$, this dynamical effect is absent.} 
In turn, the branch cut is intimately related to the presence of an irregular singularity at spatial infinity in the effective potential, together with the outgoing boundary conditions.\footnote{Recall that the irregular singularity was removed in the effective geometry approach in Sec.~\ref{sec: effectivegeometry}, such that the solution in the inner region did not display a branch cut. In contrast, the ingoing solution (dictated by the behavior at the BH horizon) displays this effect in neither the inner nor the outer region.}

Once properly taken into account in the Green's function, the brunch cut is responsible for the existence of the late-time tails. Indeed, combining the two solutions in Eq.~\eqref{Sinfty}, one can show that the contribution to the Green's function from the branch cut is~\cite{Leaver:1986gd,Andersson:1996cm}
\begin{align}
& G_\text{\tiny tail} (r_1, 0; r_2,t) \nonumber \\
& = 2 {\rm i} r_s r_1 r_2 \int_0^{- {\rm i} \infty} \d \omega \, \omega^2 j_\ell (\omega r_1) j_\ell (\omega r_2) {\rm e}^{-{\rm i} \omega t} \nonumber \\
& \simeq (-1)^{\ell+1} \frac{(2\ell+2)!}{[(2\ell+1)!!]^2} \frac{2 r_s (r_1 r_2)^{\ell+1}}{t^{2\ell+3}}\,,
\label{Gtail powerlaw}
\end{align}
where, before integrating, we have expanded the frequency-domain Green's function within the inner region,~$\omega r_{1,2} \ll 1$,  to obtain~$G_\text{\tiny tail}(r_1,r_2; \omega) \propto (r_1 r_2)^{\ell+1} \omega^{2\ell+2}$. The behavior~$G (r_1, 0; r_2,t) \sim t^{-2\ell-3}$ denotes precisely the existence of a late time tail, and its power-law does not depend on the exact nature of the central object. 

The frequency behavior of the retarded Green's function induced by tail effects shows a possible mix with the series expansion performed in Eq.~\eqref{Gana}, which assumes an instantaneous reaction, and may affect the extraction of the corresponding tidal response coefficients as done in Eq.~\eqref{GFEFT}. In a schematic decomposition, we would therefore expect the retarded Green's function to take the form
\begin{equation}
\label{instRR}
G_\text{\tiny ret} (\omega) \simeq G_\text{\tiny inst} (\omega) + G_\text{\tiny RR} (\omega)\,,
\end{equation}
where~$G_\text{\tiny inst} (\omega)$ refers to Eq.~\eqref{Gana}, and~$G_\text{\tiny RR} (\omega)$ describes  radiation reaction effects, including the late-time tail contribution~$G_\text{\tiny tail} (\omega)$. 

This issue has also been investigated in Refs.~\cite{Saketh:2023bul, Ivanov:2024sds}, when discussing the dynamical response of Kerr BHs using the formalism of scattering amplitudes. Tail effects, represented by loop corrections to scattering off the long-range Newtonian potential, give rise to logarithmic terms in the near zone and can obscure the matching of the response coefficients beyond leading order. While the EFT explicitly allows to calculate tail corrections to BH absorption and to identify them in the scattering phase,  a proper procedure to disentangle them within the general relativistic approach is still missing. 

Below we offer another perspective on the contamination between tidal response and tail effects in the Green's function, based on the Kramers-Kronig relations.

\subsection{Kramers-Kronig relations}
\label{sec: KK}
\noindent
The Kramers-Kronig relations connect the real and imaginary parts of any complex function that is analytic in the upper-half complex plane. They are commonly used to derive either the real or imaginary part of response functions for physical systems, and are based on the principle that stable systems exhibit causality, which in turn guarantees analyticity in the upper complex plane~\cite{PhysRev.104.1760}. In the context of field theory, they are intimately related to the optical theorem.

\begin{figure}[t!]
	\centering
 	\includegraphics[width=0.5\textwidth]{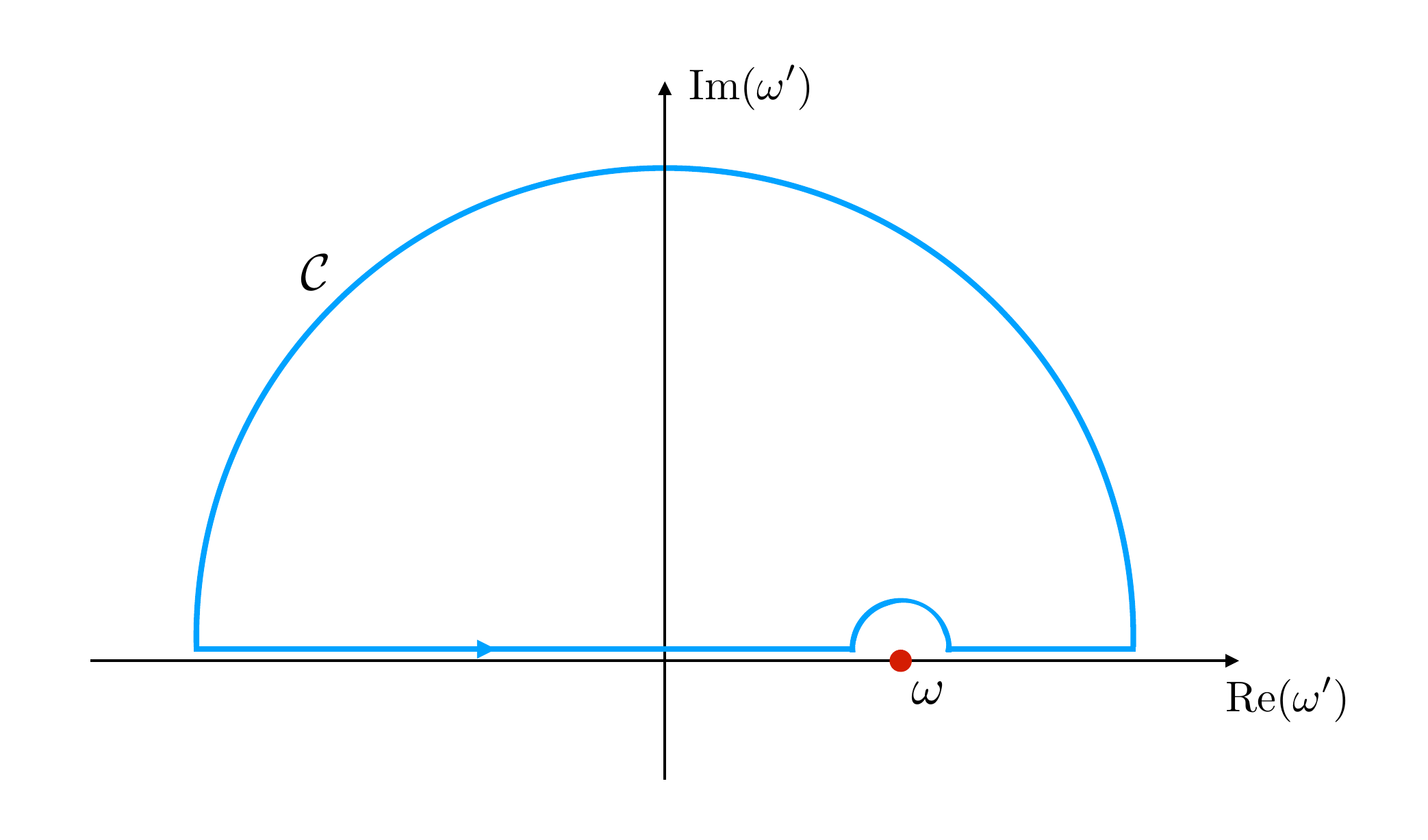}
	\caption{\it Contour integral~$\mathcal{C}$ to derive the Kramers-Kronig relations.}
	\label{fig: KK}
\end{figure}

One can apply the property of analyticity of the retarded Green's function in the upper complex frequency plane ($ {\rm Im} (\omega) > 0$), describing the response of a BH to an external perturbation, to derive corresponding relations between its real and imaginary parts. The starting point is the following integral
\begin{equation}
\mathcal{I} = \frac{1}{{\rm i}\pi} \oint_\mathcal{C} \d \omega' \frac{G(\omega')}{\omega'-\omega}\,, \quad \omega \in \mathbb{R}
\end{equation}
over the complex contour~$\mathcal{C}$, shown in Fig.~\ref{fig: KK}, which skims just above the real axis, avoiding the pole at~$\omega' = \omega$, and closing at infinity in the upper-half plane. By Cauchy's theorem, the absence of poles implies~$\mathcal{I} = 0$. 

The integral is then decomposed into three regions:~$i)$ the real axis;~$ii)$ the small semi-circle around~$\omega$; and~$iii)$ the arc at large frequencies. 
Assuming that the integrand falls off at least as~$1/|\omega'|$ at large frequencies, the large arc gives a vanishing contribution, leaving us with 
\begin{equation}
\label{KK}
G(\omega) = \frac{1}{{\rm i}\pi }\mathcal{P} \int_{-\infty}^\infty  \d \omega' \frac{G(\omega')}{\omega'-\omega}\,,
\end{equation}
where the small circle integral has been evaluated using the Sokhotski–Plemelj theorem, and where~$\mathcal{P}$ denotes the Cauchy principal value. 
Because of the~${\rm i}$ on the right-hand side, Eq.~\eqref{KK} relates the real and imaginary components of the retarded Green's function:
\begin{align}
{\rm Re} [G(\omega)] & = \frac{1}{\pi }\mathcal{P} \int_{-\infty}^\infty  \d \omega' \frac{{\rm Im} [G(\omega')]}{\omega'-\omega}\,; \nonumber \\
{\rm Im} [G(\omega)] & = - \frac{1}{\pi }\mathcal{P} \int_{-\infty}^\infty  \d \omega' \frac{{\rm Re}[G(\omega')]}{\omega'-\omega}\,. 
\end{align}
These are the well-known Kramers-Kronig relations. They follow from causality and show that the dissipative (imaginary) part of the response function is determined in terms of the conservative (real) part, and vice versa. The relationship is, however, nonlocal in frequency space, as it requires knowledge of one part of the Green's function across all frequencies to reconstruct the other for any individual frequency.

Evaluating the Green's function in the outer region ($r_{1,2} \gg r_+$), we can split the integral into buffer zone ($\omega r_{1,2} \lesssim 1$) and wave zone ($\omega r_{1,2} \gtrsim 1$) contributions (see Fig.~\ref{scales}): 
\begin{align}
G(r_1, r_2; \omega) & = \frac{1}{{\rm i}\pi }\mathcal{P} \int_{\omega' r_{1,2} \,\lesssim \,1}  \d \omega' \frac{G(r_1, r_2;\omega')}{\omega'-\omega} \nonumber \\
& + \frac{1}{{\rm i}\pi }\mathcal{P} \int_{\omega' r_{1,2}\, \gtrsim \, 1}  \d \omega' \frac{G(r_1, r_2;\omega')}{\omega'-\omega}\,,
\end{align}
This shows that knowledge of the Green's function (and tidal response) at a certain frequency requires knowledge of both near and wave zones, with the consequent relevance of tail effects. This form of the Green's function encompasses the physical information shown in Eq.~\eqref{instRR}.

Lastly, let us focus on the static limit~($\omega \to 0$), where the retarded Green's function is dominated by the leading instantaneous contribution associated to 
tidal effects. In this limit Eq.~\eqref{KK} becomes 
\begin{align}
G(r_1, r_2; 0) & \simeq G_\text{\tiny inst} (r_1, r_2; 0)  \nonumber \\ & =  \frac{1}{{\rm i}\pi }
\mathcal{P} \int_{\omega' r_{1,2} \,\lesssim\, 1}  \d \omega' \frac{G(r_1, r_2;\omega')}{\omega'}  \nonumber \\
&  +   \frac{1}{{\rm i}\pi }  \mathcal{P} \int_{\omega' r_{1,2} \,\gtrsim \, 1}  \d \omega' \frac{G(r_1, r_2;\omega')}{\omega'}\,,
\end{align}
which establishes a connection between the near and wave zones. The explicit check of such identities is left for future work, once a complete investigation of the retarded Green's function is performed.

\section{The special case: BTZ black holes}
\label{sec: BTZ}
\noindent
As discussed in the previous section, the presence of radiation reaction and tail effects can contaminate the solution of the wave equation in the buffer region and the corresponding extraction of the dynamical response from the retarded Green's function.

The purpose of this section is to study a simple case where TLNs and the Green's function can be directly related in general relativity; namely the class of BTZ black hole solutions in 2+1 dimensional gravity~\cite{Banados:1992wn, Banados:1992gq}. These solutions are special since gravity in~$2+1$ dimensions is nondynamical, hence there are no radiative modes.

Nonextremal rotating BTZ BHs are solutions to Einstein's equations with a negative cosmological constant~$\Lambda = - 1/l^2$. Their metric, in~$(t,r,\varphi)$ coordinates, reads
\begin{equation}
\d s^2 = - f(r) \d t^2 + \frac{\d r^2}{f(r)}  + r^2 \lp \d \varphi - \frac{J}{2 r} \d t \rp^2\, .
\end{equation}
The function~$f(r)$ depends on the BH mass~$M$ and angular momentum~$J$ as~\footnote{We set~$G = \frac{1}{8}$ following the standard convention.} 
\begin{equation}
f(r) = - M + \frac{r^2}{l^2} + \frac{J^2}{4 r^2} = \frac{\left(r^2-r_-^2\right) \left(r^2-r_+^2\right)}{l^2 r^2}\,,
\end{equation}
where the inner and outer BH radii are given by
\begin{equation}
r_\pm = l \left[ \frac{M}{2} \lp 1 \pm \sqrt{1 - \frac{J^2}{M^2 l^2}} \rp \right]^{1/2}\,.
\end{equation}
Consider a minimally coupled massless scalar field propagating on this background. 
Adopting the standard decomposition
\be
\Phi(t,r,\varphi) = \frac{\tilde{R}_m(r)}{\sqrt{r}}  {\rm e}^{-{\rm i} \omega t} {\rm e}^{{\rm i}m \phi}\,,
\ee
in terms of the angular eigenvalue~$m$, the radial function~$\tilde{R}_m(r)$ satisfies the Schrödinger-like equation~\eqref{Schrodinger},
with tortoise coordinate~${\rm d}x = {\rm d}r/f(r)$ given by 
\begin{equation}
x =  \frac{l^2 \left[r_- \tanh ^{-1}\left(\frac{r}{r_-}\right)-r_+ \tanh ^{-1}\left(\frac{r}{r_+}\right)\right]}{r_+^2- r_-^2}\,.
\end{equation}
Notice that the tortoise coordinate for BTZ BHs takes a finite value~$x \to \frac{{\rm i} \pi  l^2}{2 (r_+ + r_-)}$ as~$r \to \infty$.
The effective potential in Eq.~\eqref{Schrodinger} is
\begin{align}
V_m(r)  = \frac{J \omega m}{r^2} + \frac{\left(\frac{r^2}{l^2}-M\right)m^2}{r^2}  - \frac{f^2(r)}{4 r^2}-\frac{f(r) f'(r)}{2 r}\,.
\label{Veff BTZ}
\end{align}
Evidently, the potential diverges at spatial infinity (as can be seen by expanding its last term), so we will impose~$\Phi \rightarrow 0$ as~$r\rightarrow \infty$ in order to obtain a regular solution. 

A wave equation of this form can be solved in terms of hypergeometric functions. Defining~$z = \frac{r^2 - r_+^2}{r^2 - r_-^2}$, one can bring the radial equation to the form\footnote{As before,~$' = \frac{{\rm d}}{{\rm d}z}$.} 
\begin{equation}
\label{hypergeo}
z(1-z) R_m'' + (1-z) R_m' + \lp \frac{\alpha}{z} + \beta  \rp R_m = 0\,, 
\end{equation}
with  
\begin{align}
\alpha & = \frac{l^4}{4 \left(r_+^2 - r_-^2\right)^2} \lp \omega r_+ - \frac{m}{l} r_- \rp^2\,; \nonumber \\
\beta & = - \frac{l^4}{4 \left(r_+^2 - r_-^2\right)^2} \lp \omega r_- - \frac{m}{l} r_+ \rp^2 \,. 
\label{alpha beta def}
\end{align}
This equation has singular points at the inner~$z \to \infty \, (r \to r_-)$ and outer~$z \to  0 \, (r \to r_+)$ BH horizons, and at spatial infinity~$z \to 1 \, (r \to \infty)$, with two independent solutions in the neighbourhood of each singular point.

In the physically interesting range~$z \in [0,1]$, the general solution around~$z = 0$ takes the form
\begin{align}
\label{solhorgeneral}
R_m (z) & =  z^{\frac{c-1}{2}}  \Big[C_1 \, {}_2 F_1 (a,b;c;z) \nonumber \\
& + C_2 z^{1-c} {}_2 F_1 (a-c+1,b-c+1;2-c;z)\Big]\,,
\end{align}
where
\begin{align}
a, b &= - {\rm i} \sqrt{\alpha} \pm {\rm i} \sqrt{-\beta}\,; \nonumber \\
c  &= a+b+1 = - 2{\rm i} \sqrt{\alpha} + 1 \, ,
\label{abc def}
\end{align}
and the constants~$C_1,C_2$ are fixed by boundary conditions. Perturbations around BTZ BHs satisfy the boundary conditions of ingoing waves near the outer horizon~$z = 0$, and decaying at spatial infinity~$z = 1$ (as dictated by the divergent effective potential)~\cite{Horowitz:1999jd,Birmingham:2001hc}. The former condition demands that~$C_2 = 0$, such that the solution near~$z = 0$ is~\cite{Cardoso:2001hn}
\begin{equation}
\label{solhor}
R_m^+ (z) = C_1 z^{\frac{c-1}{2}} \, {}_2 F_1 (a,b;c;z)\,.
\end{equation}
This properly recovers the ingoing behavior~$R_m^+ (z) \propto (r-r_+)^{(c-1)/2}$~\cite{Natsuume:2020snz}. 

Knowledge of the ingoing solution~$R_m^+ (z)$ allows us to extract the TLNs by studying the limit~$z \to 1$ ({\it i.e.}, approaching spatial infinity). 
Since~$c-a-b = 1$ is an integer in our case, to perform the asymptotic expansion one must use a modified version of the Kummer property. Instead of Eq.~\eqref{Kummer}, the
identify is given by~\cite{abramowitz+stegun}
\begin{align}
& {}_2 F_1 (a,b;a+b+1;z) = \frac{\Gamma (a+b+1)}{\Gamma(a+1) \Gamma(b+1)} \nonumber \\
&~~~~~ \Bigg\{1 + (1-z) ab \sum_{j = 0}^\infty \frac{(a+1)_j (b+1)_j}{(j!) (j+1)!} (1-z)^j \nonumber \\
& ~~~~~~~~~~~ \times \Big[ \log (1-z) - \psi(j+1) - \psi (j+2) \nonumber \\
& ~~~~~~~~~~~~~~~~ + \psi (a+j+1) + \psi (b+j+1)\Big] \Bigg\}\,,
\end{align}
where~$(a)_j = \frac{\Gamma(a+j)}{\Gamma(a)}$ is the Pochhammer symbol and $\Gamma(z+1) = z \Gamma(z)$.
Thus one can then write the ingoing solution~\eqref{solhor} in the approximated form
\begin{align}
& R_m^+ (z) \simeq C_1 z^{\frac{a+b}{2}} \frac{\Gamma (a+b+1)}{\Gamma(a+1) \Gamma(b+1)} \nonumber \\
& ~~~~~~~~\, \times \Bigg\{1  + (1-z) ab  \Big[ \log (1-z) - 1 + 2\gamma_\text{\tiny E}\Big] \nonumber \\
&~~~~~~~~~ + (1-z) ab \Big[\psi (a+1) + \psi (b+1) + \dots \Big] \Bigg\}\,,
\label{insol2}
\end{align}
where~$\gamma_\text{\tiny E}$ is Euler's constant, and where the dots denote subleading terms in~$(1-z)$ associated to  $j \geq 1$.
Contrary to the Kerr solution, the presence of a logarithmic factor can confuse the identification of growing (source) term  and falloff (response) term. As discussed in~\cite{Hui:2020xxx, Natsuume:2020snz}, this term highlights the existence of regularization-dependent contact terms, which give rise to a classical renormalization group running of TLNs~\cite{Hui:2020xxx}. To cure this issue, one should add a proper boundary counterterm in the bulk action~\cite{Natsuume:2020snz}, which will remove them.

Since the proper identification of this contribution is not relevant for the main purpose of this work, we will follow the approach taken in~\cite{Natsuume:2020snz}, where the log term and Euler's constant are absorbed as a subleading correction to the source series. Thus, the ingoing solution~\eqref{insol2} takes the form
\begin{align} 
& R_m^+ (z) = C_1 z^{\frac{a+b}{2}} \nonumber \\
&~~~~~ \Bigg\{A \Big[ 1 + (1-z) ab  \big( \log (1-z) - 1 + 2\gamma_\text{\tiny E}\big) \Big] \nonumber \\
&  ~~~~~ + B  (1-z) \Big[1 + \cdots \Big] \Bigg\}\,,
\end{align}
where the dots indicate subleading terms. The coefficients~$A$ and~$B$ associated respectively to the growing and decaying behaviors are readily determined from~\eqref{insol2}:
\begin{align}
A &= \frac{\Gamma (a+b+1)}{\Gamma(a+1) \Gamma(b+1)}\,; \nonumber \\
B &= A \, ab\left[\psi (a+1)+ \psi (b+1)\right]\,.
\label{ABdef}
\end{align}

The tidal response is given as before as the ratio of these two coefficients:
\begin{align}
\label{TLNBTZ}
 k_m^\text{\tiny BTZ} & = a b \Big[\psi (a+1)+ \psi (b+1)\Big]\,.
\end{align}
In the low frequency limit, substituting~\eqref{alpha beta def} and~\eqref{abc def}, we obtain
\begin{widetext}
\centering
\begin{align}
k_m^\text{\tiny BTZ} & = \frac{l^2 m^2}{4 \left(r_+^2-r_-^2\right)} \left[\psi\left(\frac{{\rm i} l m}{2 (r_-+r_+)}+1\right)+ \psi\left(\frac{{\rm i} l m}{2 (r_- - r_+)}+1\right) \right] \nonumber \\
& ~~~ +\frac{{\rm i} l^4 m^2 \omega}{8 \left(r_+^2-r_-^2\right)^2}  \bigg[(r_+-r_-) \psi'\left(\frac{{\rm i} l m}{2 (r_-+r_+)}+1\right) + (r_++r_-) \psi'\left(\frac{{\rm i} l m}{2 (r_- - r_+)}+1\right)\bigg]\,,
\end{align}
\end{widetext}
\normalsize
with~$\psi'(z) = {\rm d}\psi(z)/{\rm d}z$.
Evidently, the response contains both a real and imaginary part, showing that slowly perturbed BTZ BHs have nonvanishing TLNs and experience tidal dissipation. Notice in particular that the TLNs are proportional to the cosmological constant~$l$.

What made the explicit calculation of TLNs for BTZ BHs possible is of course the simple nature of gravity in three dimensions. There are no propagating gravitational waves, nor tail effects (as discussed below), in three dimensions. The same simplifying features also allow us to easily relate the tidal response to the retarded Green's function, which as we will see captures the full response to external perturbations.

As before, to construct the Green's function we need both the ingoing solution~$R_m^+ (z)$, given by~\eqref{solhor}, as well as the outgoing solution~$R_m^\infty (z)$.
To derive the latter, we note that the hypergeometric nature of the differential equation~\eqref{hypergeo} implies the existence of a linear relation among any three solutions~\cite{abramowitz+stegun}. In particular, one can express a solution around spatial infinity,~$z = 1$, from the two solutions around~$z = 0$:
\begin{widetext}
\centering
\begin{align}
(1-z)^{c-a-b} \, {}_2 F_1 (c-a,c-b;c-a-b+1;1-z) & = \frac{\Gamma(1-c)\Gamma(c-a-b+1)}{\Gamma(1-a)\Gamma(1-b)} {}_2 F_1 (a,b;c;z) \nonumber\\
& - \frac{\Gamma(c)\Gamma(c-a-b+1)}{(1-c)\Gamma(c-a)\Gamma(c-b)} z^{1-c}{}_2 F_1 (a-c+1,b-c+1;2-c;z)\,.
\label{newkummer}
\end{align}
\end{widetext}
\normalsize
The right-hand side is a linear combination of the form given in Eq.~\eqref{solhorgeneral}, for a particular choice of~$C_1$ and~$C_2$ coefficients. In other words, up to an overall factor of~$z^{\frac{c-1}{2}}$, this is a particular solution to the hypergeometric equation~\eqref{hypergeo}. In fact, this solution falls off asymptotically (as~$z\rightarrow 1$), and thus is identified with the solution~$R_m^\infty (z)$ satisfying the outgoing boundary condition~\cite{Cardoso:2001hn}:
\begin{align}
\label{solinfty}
R_m^\infty (z) & =  K_2 z^{\frac{a+b}{2}} (1-z) \, {}_2 F_1 (a+1,b+1;2;1-z)  \,,
\end{align}
where~$K_2$ is a constant, and we have used~$c = a+b+1$. We should stress that is an {\it exact} solution to the problem. This is unlike the situation for Kerr BHs, where the inner region approximation discussed in Sec.~\ref{sec: effectivegeometry} did not allow for a hypergeometric solution at spatial infinity (technically excluded from the domain of validity of the assumption).

Equations~\eqref{solhor} and \eqref{solinfty} therefore provide the two independent solutions with which to build the retarded Green's function. Using Eq.~\eqref{newkummer}, the Wronskian of the physical solutions~$\tilde{R}^+_m =\sqrt{r} R^+_m$ and~$\tilde{R}^\infty_m = \sqrt{r} R^\infty_m$ is easily computed:\footnote{To simplify the expression we have used the following identity for the Wrosnkian between the hypergeometric functions
\begin{align}
& W\left[{}_2 F_1 (a,b;c;z), z^{1-c}  {}_2 F_1 (a-c+1,b-c+1;2-c;z)\right] \nonumber \\
&~~= (1-c) z^{-c} (1-z)^{c-a-b-1}\,.
\end{align}}
\begin{align}
W [\tilde{R}_m^\infty, \tilde{R}_m^+] &= r W [R_m^\infty, R_m^+]\nonumber \\
& = - C_1 K_2 \frac{2(r_+^2-r_-^2)}{l^2}  \frac{\Gamma (a+b+1)}{\Gamma(a+1) \Gamma (b+1)} \nonumber \\
&   = - C_1 K_2 \frac{2A(r_+^2-r_-^2)}{l^2} \,,
\label{wronskian BTZ}
\end{align}
where we have used~\eqref{ABdef} in the last step. The result is just a constant. As a check on this result, the zeros of the Wronskian
are located at the singularities of~$\Gamma(a+1)$ and~$\Gamma(b+1)$:
\begin{equation}
a+1 = - n\,; \quad b+1 = - n\,; \qquad  n \in \mathbb{N}\,.
\label{ab sing}
\end{equation}
In turn the zeros of the Wronskian coincide with the poles of the Green's function, which give the QNM frequencies.
Substituting~\eqref{alpha beta def} and~\eqref{abc def} in~\eqref{ab sing}, we find the QNM frequencies of BTZ BHs:
\begin{equation}
\omega_\text{\tiny QNM} = \pm \frac{m}{l} - \frac{2{\rm i}}{l^2} (r_+ \mp r_-) \lp n + 1 \rp\,.
\end{equation}
As expected, the imaginary part of the quasinormal frequencies satisfies the universal scaling behavior~${\rm Im}(\omega) \sim \frac{r_+}{\gamma}$, which is proportional to the BH horizon and Choptuik scaling coefficient~$\gamma = 1/2$~\cite{Govindarajan:2000vq, Birmingham:2001hc}.

Returning to the retarded Green's function, we substitute~\eqref{solhor},~\eqref{solinfty} and~\eqref{wronskian BTZ} into~\eqref{eq:GreenFunction_Definition} to obtain 
\begin{widetext}
\centering
\begin{align}
\label{GreenBTZseries}
 G_\text{\tiny BTZ} (x_1 < x_2;\omega) & = - \frac{l^2}{2 A (r_+^2-r_-^2)}  z_1^{\frac{a+b}{2}} \, {}_2 F_1 (a,b;a+b+1;z_1 )  \;  z_2^{\frac{a+b}{2}}  (1-z_2)  \, {}_2 F_1 (a+1,b+1;2;1-z_2) \nonumber \\
& \propto - \llp 1 + \frac{B}{A} (1-z_1) \rrp  (1-z_2)  \nonumber \\
& \simeq - \llp 1 + k_m^\text{\tiny BTZ} \lp \frac{r_+^2-r_-^2}{r_1^2} \rp \rrp \lp \frac{r_+^2-r_-^2}{r_2^2} \rp\,,
\end{align}
\end{widetext}
\normalsize
where in the second step we have assumed that both observer and source are situated far away from the BH,~$x_{1,2} \gg r_+$, or equivalently~$z_{1,2} \to 1$,
following the discussion of the asymptotic expansion in Sec.~\ref{sec: GF}. In the last step we have identified the tidal response from Eq.~\eqref{TLNBTZ}. 

Similarly to Sec.~\ref{sec: GF}, the first term in~\eqref{GreenBTZseries}, which is proportional to unity, describes the signal propagation directly from the source to the observer. As such, this term does not carry any information about the BH response. On the other hand, the second term, which is proportional to~$k_\text{\tiny BTZ}$, captures the full response of the object, and carries information about the TLNs, dissipative effects and the QNM frequencies. 

Let us again contrast this result with the analysis of Kerr BHs. Although the solution in the inner region of Kerr BHs is of the hypergeometric form (see~\eqref{solhorRN}), it cannot be used to build a Green's function of the form similar to Eq.~\eqref{GreenBTZseries}. This is because the outgoing solution at spatial infinity is not hypergeometric and is further characterized by tail effects.

To draw another contrast, the behavior of the BTZ effective potential~\eqref{Veff BTZ} at spatial infinity does not induce a late-time power law tail in the Green's function. This is unlike the power-law tail in asymptotically flat spacetimes, as seen in Eq.~\eqref{Gtail powerlaw}. Instead, the retarded Green's function for BTZ BHs decays exponentially~\cite{Chan:1996yk, Horowitz:1999jd}. This is compatible with the hypergeometric solutions to the wave equation, which do not show any branch cut in the complex frequency plane (see related discussion in Sec.~\ref{sec: TLN&tail}). The absence of tail effects and gravitational waves allowed us to derive an explicit, closed-form relation between TLNs and retarded Green's function.

Following the discussion in Sec.~\ref{sec: TLN&tail}, it is interesting to show how one could connect the tidal response of BTZ BHs to the singular part of the retarded Green's function. Looking at Eq.~\eqref{GreenBTZseries}, one can deduce the corresponding advanced solution simply by replacing~$z_1 \leftrightarrow z_2$. Using~\eqref{Gsingreg}, the sum and difference of the advanced and retarded Green's functions give respectively the singular and regular parts. One finds
\begin{equation}
G_\text{\tiny sing} \supset \frac{k_m^\text{\tiny BTZ}}{r_1^2 r_2^2}\,; \qquad G_\text{\tiny reg} \not \supset k_m^\text{\tiny BTZ}\,,
\end{equation}
which shows that only the singular part is proportional to the BH response coefficient~$k_m^\text{\tiny BTZ}$. This clean decomposition of the retarded Green's function into its singular and regular contributions (with~$G_\text{\tiny reg}$ not containing the TLNs) is once again a direct consequence of the absence of tail effects. This is contrast with Eq.~\eqref{Greenouter} for Schwarzschild BHs, where the presence of the logarithmic term makes the source contribution asymmetric in~$x_1 \leftrightarrow x_2$, and as such does not allow one to unambiguously isolate the source. Let us finally stress that the generalisation to the case of massive perturbations on BTZ BHs would lead to a similar decomposition of the retarded Green function, even though caution is needed when taking the massless limit to extract the TLN.

The tidal response~$k_m^\text{\tiny BTZ}$ can be similarly obtained using the AdS/CFT duality~\cite{Son:2002sd}. Rotating BTZ BHs are dual to a (1+1)-dimensional conformal field theory with conformal dimensions~$\Delta_\pm = 1 \pm 1$ and temperatures for the right(left)-moving sectors~$T_\text{\tiny R,L} = \frac{r_+ \pm r_-}{2\pi}$. It has been shown that the retarded Green's function reads~\cite{Birmingham:2001pj,Son:2002sd,Natsuume:2020snz}
\begin{align}
G_\text{\tiny CFT} &= - (\Delta_+ - \Delta_-) 4 \pi^2 T_\text{\tiny R} T_\text{\tiny L} \frac{\Gamma(a+1) \Gamma(b+1)}{\Gamma(a) \Gamma(b)} \nonumber \\
& \times \Big[\psi (a+1)+ \psi (b+1)\Big]\,.
\end{align}
The functional dependence agrees with Eq.~\eqref{GreenBTZseries}, once the radial dependence~$(r_1 r_2)^{-2}$ is properly factored out.

\section{Conclusions}
\label{conclusions}
\noindent
Response theory provides the framework to characterize the reaction of a system to external perturbations. It is usually described in terms of the retarded Green's function, as demanded by the principle of causality. In the case of interest of a compact object subject to external tidal perturbations, linear response theory allows us to describe the conservative response in terms of a set of coefficients called tidal Love numbers. These vanish identically for asymptotically flat BHs under the influence of static external perturbations, and have recently been generalized to finite frequency.

The purpose of this paper is two-fold. First, we have highlighted the intrinsic connection between dynamical response and the retarded Green's function. Second, we have demonstrated the ambiguity in extracting the former from the latter when radiation reaction and related phenomena are present.

For asymptotically flat BHs, the retarded Green's function receives three contributions:~$i)$~from high-frequency modes propagating directly from the source to the observer;~$ii)$~from quasinormal modes emitted during the relaxation to equilibrium of the perturbed BHs; and~$iii)$~from tail effects induced by the backscattering of the emitted gravitational waves on the asymptotic curved spacetime. As we have seen, the presence of tail effects, which provide an example of radiation reaction on the system induced by physics at large distances, contaminates the extraction of dynamical Love numbers. 

After reviewing the formalism of asymptotic matching, which relies on the existence of a buffer region of overlap between the inner and outer regions,
we have computed the dynamical response for Kerr BHs. Using the effective geometry approach, we have recovered the result that the conservative response of Kerr BHs
vanishes to leading order in the perturbation frequency, while the dissipative part is nonvanishing.

The extraction of TLNs within the buffer region, however, poses a problem of contamination from radiation-reaction effects. Indeed, we have shown that the retarded Green's function can be decomposed into a regular contribution, which describes the BH geodesic motion and is affected by radiation effects, and a singular contribution, which does not depend on the outgoing boundary condition at spatial infinity and contains all the relevant information on tidal effects. 

The regular contribution to the Green's function, supported by the presence of tail effects in the wave solution, contaminates the direct extraction of TLNs from the retarded Green's function. In other words, the retarded Green's function contains additional terms beyond the instantaneous contributions (from which TLNs are usually read) arising from radiation effects. As a further indication of this phenomenon, we have studied the Kramers-Kronig relations for BHs, which are dictated by the causal nature of the tidal problem, showing that near and far-zone effects are mixed together in the full response. 
A clear disentanglement is possible at least in the context of scattering amplitudes~\cite{Saketh:2023bul}. 

To provide an exception to this phenomenon, we have studied in detail the tidal response of BTZ BHs. The  absence of gravitational radiation, and hence tail effects, in 2+1 dimensions allows for a one-to-one correspondence between the instantaneous tidal response and the retarded Green's function. Correspondingly, the radial equation can be solved exactly in terms of hypergeometric functions and does not show tail effects in the solution. This allowed us to easily extract the dynamical response, to construct the retarded Green's function, and to make a direct connection between them.

This work represents an initial step toward a full investigation of the relation between tidal interactions and the retarded Green's function. It can be improved in various directions. First, it would be important to investigate alternative ways of solving the wave equation within the buffer region, based on analytical or numerical approaches, to have a more in-depth investigation of dynamical TLNs and dissipative coefficients. A promising strategy would be to consider higher-order solutions in the MST method, which has proven to be a powerful approach for solving the Teukolsky equation. Second, it would be interesting to perform a more detailed study to understand how radiation reaction and tail effects enter the solution to the wave equation, in order to possibly disentangle their contribution from the tidal one. Finally, the complete knowledge of the retarded Green's function could also allow us to explicitly check the Kramers-Kronig relations for BHs, which could provide additional insights into the interplay between the near and wave zone physics. We leave these studies for future work.

\section*{Acknowledgments}
\noindent
We thank M.~Ivanov, P.~Pani and S.~Wong for interesting discussions.
V.DL. and A.G. are supported by funds provided by the Center for Particle Cosmology at the University of Pennsylvania. 
The work of J.K. and M.T. is supported in part by the DOE (HEP) Award DE-SC0013528.

\bibliography{draft}

\end{document}